\documentclass[aps, prd,twocolumn,superscriptaddress,nofootinbib,preprintnumbers]{revtex4-1}
\usepackage[utf8]{inputenc}
\usepackage[english]{babel}
\usepackage{natbib}
\def\araa{ARA\&A}%
\def\apj{ApJ}%
\def\aap{A\&A}%
\def\mnras{MNRAS}%
\def\prd{Phys.~Rev.~D}%
\def\jcap{JCAP}%
\def\physrep{Phys.~Rep.}%
\usepackage{graphicx}
\usepackage{amsmath,amssymb}
\usepackage{hyperref}
\usepackage{braket}
\usepackage{float}
\usepackage[dvipsnames]{xcolor}
\usepackage{soul}
\usepackage{rotating}
\usepackage{multirow}
\usepackage{mathtools}
\usepackage{makecell}

\usepackage[caption = false]{subfig}
\usepackage{txfonts}
\usepackage[margin=0.75in]{geometry}
\usepackage[normalem]{ulem}

\hypersetup{
    colorlinks=true,
    linkcolor=red,
    citecolor=blue,
}

\usepackage{xcolor}

\DeclareMathAlphabet\mathbfcal{OMS}{cmsy}{b}{n}

\newcommand{\beq}{\begin{equation}}
\newcommand{\eeq}{\end{equation}}
\newcommand{\beqa}{\begin{eqnarray}}
\newcommand{\eeqa}{\end{eqnarray}}

\definecolor{darkgreen}{rgb}{0.0, 0.5, 0.0}
\definecolor{darkcyanxf}{RGB}{0.0, 139.0, 139.0}

\newcommand{\epow}[1]{\mathrm{e}^{#1}}
\newcommand{\Si}{\mathrm{Si}}
\newcommand{\Ci}{\mathrm{Ci}}

\newcommand{\planck}{\textit{Planck}}
\newcommand{\act}{\textsc{ACT}}
\newcommand{\des}{\textsc{DES}}

\newcommand{\xigty}{\mbox{$\xi_{\gamma_{t} y}$}}
\defcitealias{Battaglia:2012}{B12}

\defcitealias{paper1}{paper I}
\newcommand{\paperA}{\citetalias{paper1}}

\begin{document}
\preprint{DES-2021-0636}
\preprint{FERMILAB-PUB-21-331-AE}
\title[Shear-y]{Cross-correlation of DES Y3 lensing  and \textsc{ACT}/\textit{Planck} thermal Sunyaev Zel'dovich Effect II: Modeling and constraints on halo pressure profiles}

\label{firstpage}

\begin{abstract}
Hot, ionized gas leaves an imprint on the cosmic microwave background via the thermal Sunyaev Zel'dovich (tSZ) effect.  The cross-correlation of gravitational lensing (which traces the projected mass) with the tSZ effect (which traces the projected gas pressure) is a powerful probe of the thermal state of ionized baryons throughout the Universe, and is sensitive to effects such as baryonic feedback.  
In a companion paper (Gatti et al. 2021), we present tomographic measurements and validation tests of the cross-correlation between galaxy shear measurements from the first three years of observations of the Dark Energy Survey, and tSZ measurements from a combination of Atacama Cosmology Telescope and $\planck$ observations.  In this work, we use the same measurements to constrain models for the pressure profiles of halos across a wide range of halo mass and redshift.
We find evidence for reduced pressure in low mass halos, consistent with predictions for the effects of feedback from active galactic nuclei.  We infer the hydrostatic mass bias ($B \equiv M_{500c}/M_{\rm SZ}$) from our measurements, finding $B = 1.8\pm0.1$ 
when adopting the $\planck$-preferred cosmological parameters. We additionally find that our measurements are consistent with a non-zero redshift evolution of $B$, with the correct sign and sufficient magnitude to explain the mass bias necessary to reconcile cluster count measurements with the $\planck$-preferred cosmology.
Our analysis introduces a model for the impact of intrinsic alignments (IA) of galaxy shapes on the shear-tSZ correlation. We show that IA can have a significant impact on these correlations at current noise levels.
\end{abstract}

% Author list file generated with: mkauthlist 1.2.4 
% mkauthlist -f -d -j prd /Users/shivam/Dropbox/Downloads/DES-2019-0458_author_list_ascii.csv DES-2019-0458_author_list.tex 

% \documentclass[reprint,superscriptaddress]{revtex4-1}
% \pagestyle{empty}
% \begin{document}
% \title{DES Publication Title}

\author{S.~Pandey}\email{shivamp@sas.upenn.edu}
\affiliation{Department of Physics and Astronomy, University of Pennsylvania, Philadelphia, PA 19104, USA}
\author{M.~Gatti}
\affiliation{Department of Physics and Astronomy, University of Pennsylvania, Philadelphia, PA 19104, USA}
\author{E.~Baxter}
\affiliation{Institute for Astronomy, University of Hawai'i, 2680 Woodlawn Drive, Honolulu, HI 96822, USA}
\author{J.~C.~Hill}
\affiliation{Department of Physics, Columbia University, New York, NY, USA 10027}
\affiliation{Center for Computational Astrophysics, Flatiron Institute, New York, NY, USA 10010}
\author{X.~Fang}
\affiliation{Department of Astronomy/Steward Observatory, University of Arizona, 933 North Cherry Avenue, Tucson, AZ 85721-0065, USA}
\author{C.~Doux}
\affiliation{Department of Physics and Astronomy, University of Pennsylvania, Philadelphia, PA 19104, USA}
\author{G.~Giannini}
\affiliation{Institut de F\'{\i}sica d'Altes Energies (IFAE), The Barcelona Institute of Science and Technology, Campus UAB, 08193 Bellaterra (Barcelona) Spain}
\author{M.~Raveri}
\affiliation{Department of Physics and Astronomy, University of Pennsylvania, Philadelphia, PA 19104, USA}
\author{J.~DeRose}
\affiliation{Lawrence Berkeley National Laboratory, 1 Cyclotron Road, Berkeley, CA 94720, USA}
\author{H.~Huang}
\affiliation{Department of Physics, University of Arizona, Tucson, AZ 85721, USA}
% ...

\collaboration{DES and ACT Collaboration}

% \begin{abstract}
% This is a sample document created by \texttt{mkauthlist v1.2.4}.
% \end{abstract}
% \maketitle
% \end{document}

\maketitle

\section{Introduction}
The distribution and energetics of baryons within dark matter halos are significantly impacted by astrophysical feedback processes.  In particular, large-scale winds driven by the supernova and active galactic nuclei (AGN) are expected to reduce the ability of gas in halos to form stars, and are therefore important ingredients in our understanding of galaxy formation \citep[for a review see, e.g.,][]{Benson:2010}.  At large halo mass, feedback from AGN is expected to dominate over other feedback mechanisms \citep{Booth:2013}.  Indeed, AGN feedback is sufficiently powerful that it modifies the total matter power spectrum at wavenumbers $k \gtrsim 0.1\,h/{\rm Mpc}$ \citep{vanDaalen:2011, Schneider_2016}.  Unfortunately, because feedback effects span a wide dynamical range --- from sub-parsec scales to the scales of galaxy clusters --- they are difficult to model and simulate \citep{Somerville:2015}.  As a result, attempts to extract cosmological information from the matter power spectrum at small scales (e.g., with weak lensing surveys) are often limited by our ignorance of feedback \citep[e.g.][]{Huang:2019, Chisari:2019}.  Therefore, tighter observational constraints on feedback are of prime importance for our understanding of both galaxy formation and cosmology.  

Because feedback changes the thermal energy and distribution of the baryons, it can change the pressure of ionized gas within halos, resulting in an observable signature in the thermal Sunyaev-Zel'dovich (tSZ) effect \citep[e.g., ][]{LeBrun:2014, McCarthy:2014, Battaglia:2012, Villaescusa-Navarro:2021}.  The tSZ results from inverse Compton scattering of CMB photons with hot electrons, and the amplitude of the effect --- typically expressed in terms of the Compton $y$ parameter --- is directly sensitive to a line-of-sight integral of the ionized gas pressure \citep{Sunyaev:1972}.  However, because the tSZ effect is sensitive to the pressure of all ionized gas along the line of sight to the last scattering surface, it is difficult to use the tSZ by itself to probe the halo mass or redshift dependence of the halo gas pressure.  

By {\it cross-correlating} $y$ maps obtained from CMB observations with tracers of large scale structure observed at low redshift, contributions to $y$ from particular subsets of halos can be isolated.  Such cross-correlations therefore enable measurement of the evolution  of the pressure of ionized gas over cosmic time \citep[e.g.,][]{Vikram2017,Hill:2018,Pandey:2019,Pandey:2019b,chiang2020thermal}.  

The impact of feedback on halo pressure profiles is a function of halo mass and redshift.  At large halo mass, the energy released by feedback is small compared to the gravitational potential energy of the halo, so the impact of feedback is generally less pronounced; at low halo mass, the reverse is true.  For low-mass halos, feedback can push out a significant amount of gas from the halo, resulting in reduced pressure relative to expectations from self-similar models \citep{LeBrun:2017}. Feedback is also expected to generate significant non-thermal pressure support in low-mass halos, lowering the temperature needed to maintain equilibrium.  Redshift evolution of the pressure profiles of halos is expected for several reasons, including evolving non-thermal pressure support and the fact that at fixed halo mass, halos at high redshift have deeper potential wells, making it more difficult for feedback to expel gas \citep{LeBrun:2017}.

Here we consider the cross-correlation of the gravitational shearing of galaxy shapes with maps of the tSZ effect.  As we show below (and as was pointed out previously by \cite{Osato:2018, Battaglia:2015, Hojjati_2015}), this correlation is predominantly sensitive to the pressure profiles of halos with masses $M_{200c} \sim 10^{14} M_{\odot}$ and $z \lesssim 1$.\footnote{We use $M_{\Delta c}$ to represent the mass enclosed in a sphere centered on the halo with radius chosen such that the mean enclosed density is $\Delta \rho_{\rm crit}(z)$, where $\rho_{\rm crit}(z)$ is the critical density of the Universe at the redshift of the halo.}  One of the appealing features of the lensing-tSZ correlation is that --- unlike the galaxy-tSZ correlation --- it can be modeled without needing to understand the galaxy-halo connection.  
Several recent studies have measured the lensing-tSZ correlation \citep{vanWaerbeke:2014, Hill-Spergel2014, Hojjati2017, Osato:2018, Osato:2020}.

In this work and in a companion paper (\citet{paper1}, hereafter paper I), we present measurements and analysis of the correlation between lensing shear measurements from Year 3 observations of the Dark Energy Survey (DES) and tSZ measurements from the Atacama Cosmology Telescope (ACT) and $\planck$.  DES is a six-year optical and near-infrared galaxy survey of 5000 sq. deg. of the southern sky.  
 
ACT is a submillimeter telescope located in the Atacama desert that is currently performing the Advanced ACT survey.
We use the data collected from its ACTPol receiver during 2014 and 2015.
We detect the correlation between lensing and the tSZ at 21$\sigma$ statistical significance, the highest signal to noise measurement of this correlation to date.  

A companion paper, \paperA, presents the cross-correlation measurements, subjecting them to various systematic tests, and presents a comparison of the measurements to predictions from hydrodynamical simulations.  Here, we focus on fitting the measurements with parameterized models to explore how the halo pressure profiles vary as a function of halo mass and redshift.  We present constraints on the parameters of these models and on the inferred relationship between halo mass and the integrated tSZ signal.  Our constraints exhibit a departure from the expectations of self-similar models at low halo mass ($M \lesssim 10^{14}\, M_{\odot}$), consistent with expectations from the impact of feedback from AGN. 
We translate our measurements into constraints on the so-called mass bias parameter, finding a preference for its evolution with redshift. Such redshift evolution helps to explain the mass bias values needed to reconcile cluster abundance measurements with the cosmological model preferred by $\planck$ \cite{Planck:2018cosmo}. 
Additionally, we show that the impact of intrinsic alignments of galaxy shapes on the shear-tSZ correlation --- an effect that has been ignored in previous analyses --- can be significant, especially at low redshift.

The paper is organized as follows.  In \S\ref{sec:meas_mod} we describe the shear-tSZ correlation measurements and the various models we use to fit these; in \S\ref{sec:settings} we describe our methodology for fitting the data, including choices of parameter priors; we present our results in \S\ref{sec:results} and conclude in \S\ref{sec:conclusion}.

\section{Measurements and Modeling}
\label{sec:meas_mod}

\subsection{Measurements of the shear-\textit{y} correlations}
\label{sec:measurement}

We analyze the cross-correlation between measurements of galaxy shear from DES Y3 observations \citep{y3-shapecatalog, y3-gold} and Compton-$y$ maps generated by $\act$ \citep{Madhavacheril:2020} and $\planck$ \citep{Planck:2015comptony}.   The details of the measurement process and tests of robustness to various systematics are described in detail in \paperA.  We summarize below the key aspects of the data and measurements relevant to the present analysis.

We use the shear catalog of the DES Y3 data as presented in \citet{y3-shapecatalog}. The shape catalog primarily uses the \texttt{metacalibration} algorithm and additionally incorporates improvements in the PSF estimates \citep{y3-piff} and improved astrometric methods \citep{y3-gold}. However, this pipeline does not capture the object blending effects and shear-dependent detection biases; hence image simulations are used to calibrate this bias as detailed in \citet{y3-imagesims}. This catalog consists of approximately 100 million galaxies with effective number density of $n_{\rm eff} = 5.6$ galaxies per ${\rm arcmin}^2$ and an effective shape noise of $\sigma_{\rm e} = 0.26$. 

The source galaxy sample is divided into four tomographic bins with redshift edges of the bins equal to [0.0, 0.358, 0.631, 0.872, 2.0]. The description of the tomographic bins of source samples and the methodology for calibrating their photometric redshift distributions are summarized in \citet*{y3-sompz}. The redshift calibration methodology involves the use of self-organizing maps (SOMPZ) \citep{y3-sompz} which leverage additional photometric bands in the DES deep-field observations \citep{y3-deepfields} and the \textsc{BALROG}\ simulation software of \citet{y3-balrog} to characterize a mapping between color space and redshifts. The clustering redshift method is also used to provide additional redshift information in \citet{y3-sourcewz}.  That work uses the information in the cross-correlation of the source galaxy sample with the spectroscopic data from Baryon Acoustic Oscillation Survey (BOSS) and its extension (eBOSS). Using a combination of SOMPZ and clustering redshifts, candidate source redshift distributions are drawn and provide us with the mean redshift distribution of the source galaxies and uncertainty in this distribution.

We use two $y$ maps in this analysis, one generated from a combination of $\act$ and $\planck$ data (described in \cite{Madhavacheril:2020}) and one using $\planck$ data alone. For simplicity, we refer to these as the $\act$ and $\planck$ $y$-maps, respectively.
We construct the $\planck$ Compton-$y$ map using all the publicly available 2015 $\planck$ High Frequency Instrument (HFI) and Low Frequency Instrument (LFI) frequency maps below 800~GHz \citep{Planck:lfi_maps,Planck:hfi_maps}. We use the map generated by the constrained Needlet Internal Linear Combination (NILC) algorithm \citep{Delabrouille:2009, Remazeilles_2010}, which estimates the minimum variance Compton-$y$ map as a linear combination of the temperature maps while imposing a unit-response to the frequency dependence of Compton-$y$ and a null-response to the frequency dependence of Cosmic Infrared Background (CIB). The measurements and analysis of the cross-correlations of NILC $y$ map with other large scale structure (LSS) tracers, as studied here, largely removes the leakage of foreground to the measurements. The details of the implementation of this algorithm to obtain CIB de-projected $y$-maps used in this work are presented in Appendix A of \citet{Pandey:2019}.

The $\act$ $y$-map covers only the \texttt{D56} region, amounting to 456 square degrees of overlap with the DES shear catalog, while the $\planck$ $y$-map covers the full sky. Owing to the higher resolution and sensitivity of the $\act$ $y$ map, we only use the $\planck$ $y$-map over the region of the sky covered by DES, but not covered by the $\act$ map.  

We measure two-point correlations between the galaxy shears and Compton $y$ as a function of the angular separation of the two points being correlated.  When measuring the correlations, we consider only the component of the spin-2 shear field orthogonal to the line connecting the two points being correlated, i.e., the tangential shear $\gamma_{t}$.  
The $y$-$\gamma_t$ correlation, which we represent with $\xi_{\gamma_t,y}(\theta)$, is expected to contain all of the physical signal 
while being robust to additive systematics in the shear field.  An added advantage of this quantity is that it can be computed using the shear field directly, without constructing a lensing convergence map from the shear catalog. In \paperA, the measurements are further validated against the systematics effects of the radio sources and also shows that the cross-component of the lensing signal around the tSZ maps passes the null test.

\begin{figure*}
\centering
\includegraphics[width=\textwidth]{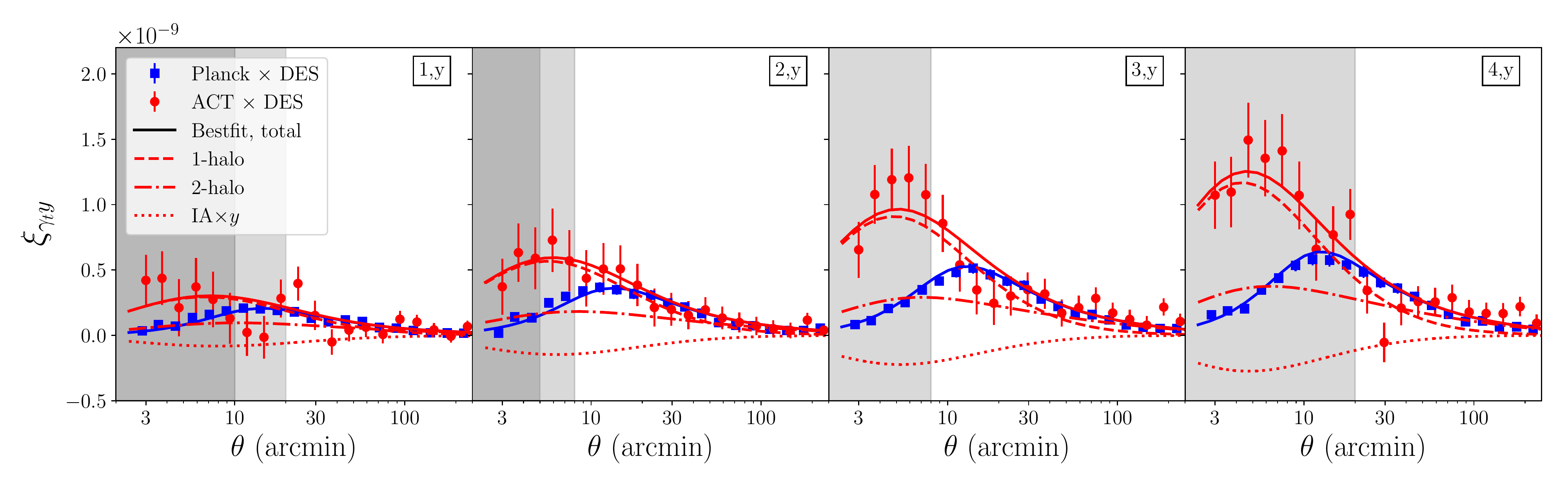}
\caption[]{Measurements of our observable, $\xi_{y\gamma_t}$, using the DES Y3 shear catalog split into four tomographic bins and Compton-$y$ map from Planck and ACT (see \paperA $\,$ for details). The shaded regions denote our scale cuts and are excluded in this analysis as they receive contributions from the cosmic infrared background and higher-order intrinsic alignment than our fiducial model. The light shaded region corresponds to the scale cuts for the $\planck \times $DES, and the dark region corresponds to the $\act \times $DES datavectors, respectively. We show the total best-fit using solid lines for both the datavectors as well using the model detailed in \S\ref{sec:meas_mod}. This total best-fit is decomposed into 1-halo, 2-halo, and intrinsic alignment (IA) correlations that are depicted using dashed, dot-dashed and dotted lines respectively for $\act \times $DES datavector. Note that the $\planck$ and the ACT Compton-$y$ maps have different beam sizes which impact the measurements in the small scales and we forward model the impact of beam in our theory model.
}
\label{fig:xigty_measure}
\end{figure*}

The final tomographic measurements of $\xi_{\gamma_t y}$ using both the $\planck$ and $\act$ Compton-$y$ maps are shown in Fig.~\ref{fig:xigty_measure}.  The correlation is detected at $21\sigma$ across all bins. The error bars correspond to the covariance estimated using a theory model (see \S\ref{sec:covariance}) and accounts for non-Gaussian sources of noise. Note that the difference in the correlations measured using the $\planck$ and $\act$ Compton-$y$ maps are due to different beam sizes of the instruments which we account for in our theory model (see \S\ref{sec:model}). We show the best fit curves obtained using our halo model framework, including contributions from intra-halo (1-halo), inter-halo (2-halo) and correlations between the intrinsic alignment of the source galaxies and Compton-$y$ (IA$\times y$). The shaded regions correspond to angular scales that are not included in our fits (note that they are different for the $\planck$ and $\act$ Compton-$y$ map correlations). These scales are excluded in order to reduce the biases from the non-linear intrinsic alignment of source galaxies and other effects at small scales that we do not include in our model (see further discussion in \S\ref{sec:settings}).

\subsection{Halo model for the shear-\textit{y} correlations}\label{sec:model}

Owing to decreasing signal-to-noise at very large angular scales and possible large-scale systematics, we restrict our analysis to scales below 250 arcminutes. Note that the shear catalog used in this analysis has been thoroughly validated for correlation analyses below 250 arcminutes \citep{Gatti:2020} and is used for cosmological analysis for scales below these scales in \citet{y3-cosmicshear1, y3-cosmicshear2}. For simplicity, then, we adopt a flat sky approximation.  In this case, the two-point angular correlation, $\xi_{\gamma_t y}(\theta)$, between galaxy shears in tomographic bin $i$, and Compton-$y$ can be related to the angular cross-power spectrum, $C_{\kappa y}(\ell)$, between the lensing convergence, $\kappa^{i}$, and Compton-$y$ via:  
\beq\label{eq:xigty}
\xi^{ij}_{\gamma_t y}(\theta) = \int \frac{d\ell \ \ell}{2\pi} J_{2}(\ell \theta) C^{ij}_{\kappa y}(\ell),
\eeq
where $J_{2}$ is the second-order Bessel function.  Here, $j$ labels the $y$ map (i.e. either $\planck$ or   $\act$), and $i$ labels the redshift bin of the galaxy lensing measurements. 

We model $C^{ij}_{\kappa y}(\ell)$ using a halo-model framework.  We will initially keep our discussion quite general, as the same modeling framework can be used (with small adjustments) to describe all of the cross-spectra needed to build our final model.  We use $A$ and $B$ to denote two tracers of the large scale structure, for instance, lensing and Compton-$y$.

In the halo model \citep[for a review see][]{Cooray:2002}, the cross-power between $A$ and $B$ can be written as the sum of a one-halo term %($C^{ij}_{\kappa y;\textrm{1h}}(\ell)$) 
and a two-halo term.
The one-halo term is given by an integral over redshift ($z$) and halo mass ($M$):

\begin{equation}\label{eq:Cl1h}
C^{i j}_{A B; \textrm{1h}}(\ell) = \int_{z_{\rm{min}}}^{z_{\rm{max}}} dz \frac{dV}{dz d\Omega} \int_{M_{\rm{min}}}^{M_{\rm{max}}} dM \frac{dn}{dM} \bar{u}^{i}_{A}(\ell,M,z) \ \bar{u}^{j}_{B}(\ell,M,z),
\end{equation}
where $dV$ is the cosmological volume element, $d\Omega$ is the solid angle constructed by that element and $dn/dM$ is the halo mass function which we model using the \citet{Tinker:2008} fitting function. In the following sub-sections we will describe the modeling of the multipole-space kernels, $\bar{u}^{i}_{A}(\ell,M,z)$ and $\bar{u}^{j}_{B}(\ell,M,z)$ of various LSS tracers. In particular we describe in detail the modeling of the lensing profile (through the  convergence field, $\kappa$) 
and intrinsic alignment ($\textrm{I}$) for any tomographic bin $i$ as well as Compton-$y$. We find that using $M_{\rm{min}}=10^{10} \, M_{\odot}/h$, $M_{\rm{max}}=10^{17} \, M_{\odot}/h$, $z_{\rm{min}}=10^{-2}$ and $z_{\rm{max}}=3.0$ ensure that the above integrals are converged.

The two-halo term, which corresponds to the inter-halo contribution to the cross-correlation, is given by:
\begin{equation}\label{eq:Cl2h}
C^{ij}_{AB;\textrm{2h}}(\ell) = \int_{z_{\rm{min}}}^{z_{\rm{max}}} dz \frac{dV}{dz d\Omega} b_{A}^{i}(\ell,z) \ b^{j}_{B}(\ell,z) \ P_{\rm{lin}}(k,z),
\end{equation}
where $P_{\rm lin}(k,z)$ is the linear matter power spectrum and $k=(\ell + 1/2)/\chi$. The terms $b_{A}^{i}(\ell,z)$ and $b^{j}_{B}(\ell,z)$ are the effective linear bias parameters describing the clustering of tracers $A$ and $B$ respectively. 
In our case, there are three tracers of interest: lensing, $y$, and intrinsic alignments.  We describe our models for these tracers in more detail below.

\subsection{Pressure profile models}
\label{sec:model_y}
The multipole-space kernel of Compton-$y$ is related to the pressure profile of hot electrons ($P_e$) as follows:
\begin{multline}\label{eq:uyl}
\bar{u}^{j}_{y}(\ell,M,z) = b^{j}(\ell) \, \frac{4\pi r_{200c}}{l^2_{200c}} \frac{\sigma_T}{m_e c^2} \int_{x_{\rm{min}}}^{x_{\rm{max}}} dx \ x^2 \ P_e(x|M,z) \\ \times \frac{\sin(\ell x/l_{200c})}{\ell x/l_{200c}},
\end{multline}
where $x = r/r_{\rm 200c}$, $r$ is the radial distance; $l_{200c} = D_A/r_{\rm 200c}$, $D_A$ is the angular diameter distance to redshift $z$ and $r_{\rm 200c}$ denotes the radius of the sphere having total enclosed mean density equal to 200 times the \textit{critical} density of the universe \citep{Komatsu_2002}.
The term $b^{j}(\ell) = \exp{[-\ell(\ell + 1) \sigma_j^2/2]}$ captures the beam of experiment $j$. Here $\sigma_j = \theta_{j}^{\rm FWHM}/\sqrt{8\ln 2}$ and we have $\theta_{1}^{\rm FWHM} = 10$ arcmin for $\planck$ and $\theta_{2}^{\rm FWHM} = 1.6$ arcmin for $\act$ Compton-$y$ maps.\footnote{Note that the full ACT beams, including variations with observing seasons season and telescope arrays have been taken into account when creating the Compton-$y$ map as described in \citet{Madhavacheril:2020}, and only the final $y$-map is reconvolved with a simple Gaussian beam.} We choose $x_{\rm{min}} = 10^{-3}$ and $x_{\rm{max}} = 4$, which ensures that the above integral captures the contribution to the pressure from the extended profile of hot gas. We have verified that our conclusions remain unchanged when lowering the value of $x_{\rm{max}}$. We have also verified that inclusion of the pixel window function of Compton-$y$ maps has negligible impact on the theory predictions as the scales analyzed to obtain our results here are significantly larger compared to the pixel size of the maps. 

The effective tSZ bias $b^j_{y}$ is given by:
\beqa\label{eq:by}\label{eq:byl}
b^j_{y}(\ell,z) = \int_{M_{\rm{min}}}^{M_{\rm{max}}} dM \ \frac{dn}{dM} \bar{u}^j_{y}(\ell,M,z) b_{\rm{lin}}(M,z),
\eeqa
where $b_{\rm{lin}}$ is the linear bias of halos with mass $M$ at redshift $z$ which we model using the \citet{Tinker:2010} fitting function.

One of the aims of this analysis is to constrain the pressure profiles of halos as a function of mass and redshift.  We consider several possible pressure profile models: one based on \citetalias{Battaglia:2012}, a modified version of this profile that allows for additional freedom to capture the impact of feedback in low-mass halos, and the model from \citet{Arnaud:2010}.  We describe each of these models in more detail below.

{\bf Battaglia et al. 2012 profile model:}  For a fully ionized gas, the total electron pressure $P^{\rm B12}_e$ that contributes to the Compton-$y$ signal is related to total thermal pressure ($P^{\rm B12}_{\rm th}$) as:
\begin{equation}
    P^{\rm B12}_e = \Bigg[ \frac{4 - 2Y}{8 - 5Y} \Bigg] P^{\rm B12}_{\rm th},
\end{equation}
where $Y$ is the primordial helium fraction that we fix to $Y=0.24$. The total thermal pressure profile in \citetalias{Battaglia:2012} is parametrized by a generalized NFW form:
\beq\label{eq:B12_model}
P^{\rm B12}_{\rm th}(x|M,z) = P_{\Delta} \tilde{P}_0 \bigg( \frac{x}{\tilde{x}_c} \bigg)^{\tilde{\gamma}} \big[ 1 + (x/\tilde{x}_c)^{\tilde{\lambda}} \big]^{-\tilde{\beta}},  
\eeq
where
\begin{equation}
P_{\Delta} = \frac{G \Delta M_{\Delta} \rho_{c}(z) \Omega_b}{2 R_{\Delta} \Omega_m},
\end{equation}
for any spherical overdensity, $\Delta$, relative to the critical density, $\rho_{c}$, and we will use $\Delta=200$. Following \citetalias{Battaglia:2012}, we fix $\tilde{\lambda} = 1.0$ and $\tilde{\gamma} = -0.3$.  For each of the parameters $\tilde{P}_0$, $\tilde{x}_c$ and $\tilde{\beta}$, \citetalias{Battaglia:2012} adopts a scaling relation with mass and redshift. This scaling relation is given by the following form (shown here for the parameter $\tilde{P}_0$):
\beq\label{eq:B12_scaling}
\tilde{P}_0(M_{200},z) = P_0 \bigg(\frac{M_{200c}}{M_{{*}}}\bigg)^{\alpha_{\rm m}} (1 + z)^{\alpha_z},
\eeq
where $P_0$ is the amplitude of the pressure profile at $M_{200c} = M_{*} \equiv 10^{14} M_{\odot}/h$ and $z=0$, and $\alpha_{\rm m}$ and $\alpha_z$ describe the scaling of the parameter $\tilde{P}_0$ with mass and redshift, respectively. Similar equations can be written down for the parameters $\tilde{x}_c$ and $\tilde{\beta}$ (with their respective mass and redshift power-law indices). We have experimented with changing the value of the break mass $M_{*}$, but find that our results are not very sensitive to this choice.  The pressure profile parameters that are not varied are fixed to the values from Table~1 of \citetalias{Battaglia:2012}.

\begin{figure*}[ht]
\centering
\includegraphics[width=\textwidth]{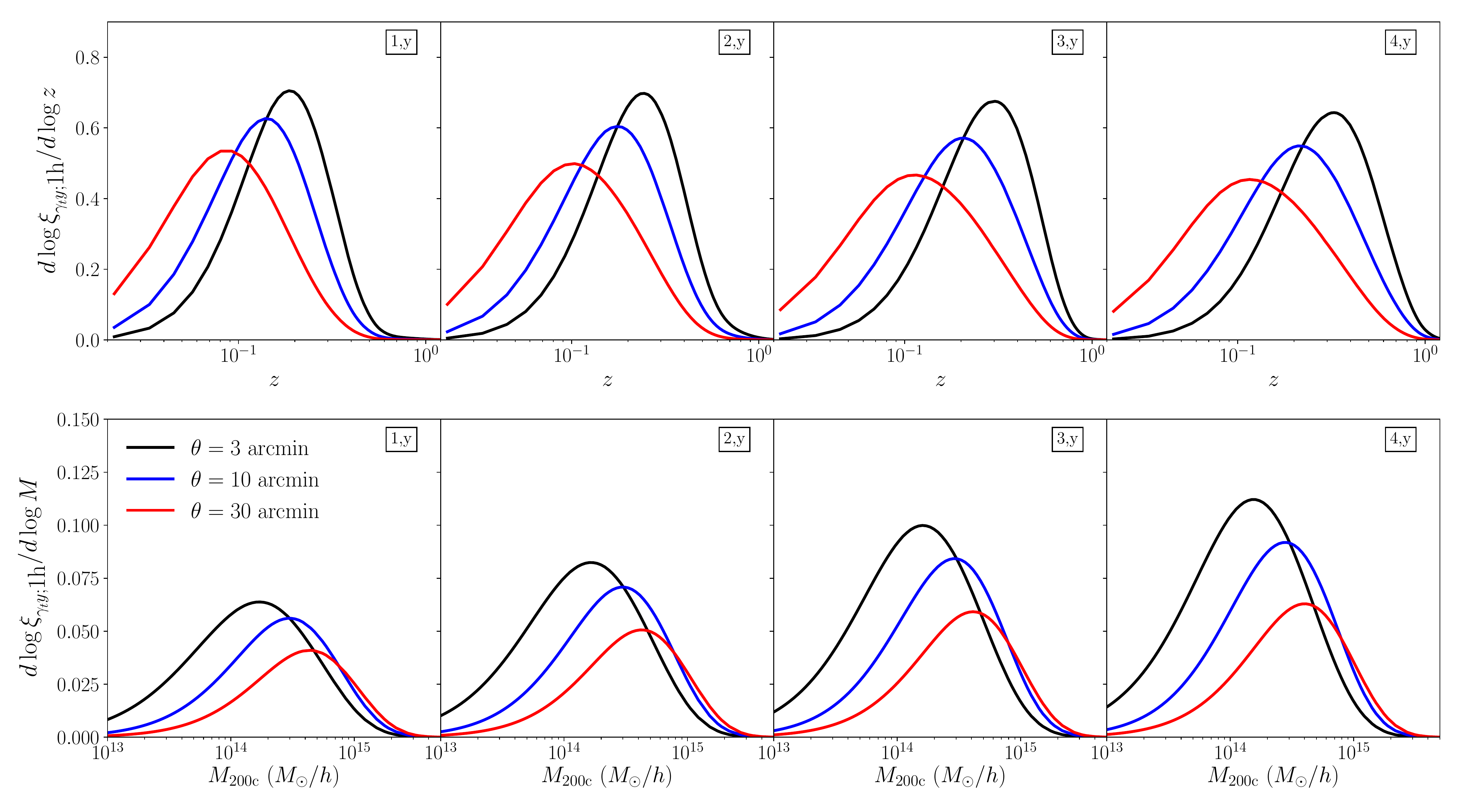}
\caption[]{Sensitivity of the one-halo contribution to the shear-$y$ correlation, $\xi_{\gamma_t y}$.  We show the logarithmic derivative of the correlation with respect to halo redshift (top) and halo mass (bottom). Note that no Compton-$y$ map beam smoothing is applied when producing these curves. The different columns represent the different redshift bins of the shear sample. To obtain this plot, we use the \textit{break model} of pressure profile (as described in \S\ref{sec:model_y}) and the parameter values of the full model are given in Table \ref{tab:params_all}. 
}
\label{fig:gty_sens}
\end{figure*}

{\bf \textit{Break model}:} The $\kappa-y$ cross-correlations receive contributions from a very wide range of halo masses (as shown in Fig.~\ref{fig:gty_sens} and discussed in \S\ref{sec:2pt_model_final}).  At low halo mass, the pressure profiles of halos may depart from the \citetalias{Battaglia:2012} form as a result of, for example, baryonic feedback.  We introduce additional freedom into our model to allow for this possibility using the formalism described in \citet{Pandey:2019b}.  We consider a modified version of the $P^{\rm B12}_e$ profile:
\beqa\label{eq:Pe_total}
P^{\rm B12, break}_e(r|M,z) =
\begin{cases}
P^{\rm B12}_e(r|M,z) \,, \ \ \ M \geq M_{\rm break} \\
P^{\rm B12}_e(r|M,z) \left( \frac{M}{M_{\rm break}} \right)^{\alpha^{\rm break}_{\rm m}},  M < M_{\rm break} 
\, \nonumber
\end{cases}
\\
\label{eq:UB}
\eeqa
where we choose $M_{\rm break} = 2 \times 10^{14} M_{\odot}/h$ and we will treat the power-law index $\alpha^{\rm break}_{\rm m}$ as a free parameter.  The location of the break is motivated by the results of simulations \citep[e.g.,][]{LeBrun:2017}, which show a break in the self-similar scaling of integrated $y$ with mass at roughly this mass value.

{\bf Arnaud et al. profile model:} We also test the \citet{Arnaud:2010} profile (denoted with A10), which is another universal profile form where its parameters have been calibrated using X-ray and tSZ observations of clusters. We note that the parameter values obtained by \citet{Arnaud:2010} are from an analysis of high mass and low redshift clusters.  The shear-$y$ correlation will be sensitive to somewhat different halos. Another crucial assumption adopted in the model of \citet{Arnaud:2010} is that the clusters are in hydrostatic equilibrium (HSE), allowing for an estimate of HSE mass. However, significant non-thermal pressure support would violate this assumption. Hence, the HSE mass can be different from the true mass of the halos. The relation between these two can by parameterized by a mass bias parameter $B$.

The \citet{Arnaud:2010} profile is:
\begin{multline}\label{eq:A10_model}
P^{\rm A10}_e(x|M,z) = 1.65 (h/0.7)^2 {\rm eV cm^{-3}} \\ \times E^{8/3}(z) \Bigg[ \frac{M^{\rm SZ}_{\rm 500c}}{3 \times 10^{14} (0.7/h) M_{\odot}} \Bigg]^{2/3 + \alpha^{\rm A10}_p} p^{\rm A10}(x),
\end{multline}
where $E(z) = H(z)/H_0$ and the generalized NFW profile $p^{\rm A10}(x)$ is given by:

\begin{equation}
    p^{\rm A10}(x) = \frac{P^{\rm A10}_0 (0.7/h)^{3/2}}{(c^{\rm A10}_{500}x)^{\gamma^{\rm A10}} \big[1 + (c^{\rm A10}_{500}x)^{\alpha^{\rm A10}}\big]^{(\beta^{\rm A10}-\gamma^{\rm A10})/\alpha^{\rm A10}} }
\end{equation}

We adopt the best-fit values obtained from the analysis of the stacked pressure profile of $\planck$ tSZ clusters, $P^{\rm A10} = 6.41$, $c^{\rm A10}_{500} = 1.81$, $\alpha^{\rm A10} = 1.33$, $\beta^{\rm A10} = 4.13$ and $\gamma^{\rm A10} = 0.31$ \citep{Planck_2013_clusters}. We also fix the parameter $\alpha^{\rm A10}_p = 0.12$ as obtained by \citet{Arnaud:2010} in their X-ray sample analysis. The mass obtained from the mass-pressure relation in Eq.~\ref{eq:A10_model} is related to the true mass of halos by the mass bias parameter, $B$. We consider a model with a constant mass bias parameter, where the true cluster mass $M_{500c}$ is related to the tSZ mass used in  Eq.~\ref{eq:A10_model} by $M^{\rm SZ}_{500c} = M_{500c}/B$ and $r_{200c}$ in Eq.\ref{eq:uyl} is replaced by $r^{\rm SZ}_{200c} = r_{200c}/(B^{1/3})$. We refer to this model as $P^{\rm A10c}_e$. We also test another model, $P^{\rm A10z}_e$, where the mass bias evolves with redshift as: 
\begin{equation}\label{eq:Bz}
    B(z) = B(1+z)^{\rho_B}
\end{equation}
We treat $B$ and $\rho_B$ as free parameters in this model. We refer the reader to Table~\ref{tab:params_all} for a concise summary of the models and their notations.

\subsection{Lensing model}
The effective multipole-space kernel of convergence can be related to the dark-matter kernel ($u_{\rm{m}}$) as: 
\begin{equation}\label{eq:ukl}
    \bar{u}_{\kappa}^{i}(\ell,M,z) = \frac{W^{i}_{\kappa}(z)}{\chi^2}  u_{\rm{m}}(k,M),
\end{equation}
where $k = (\ell + 1/2) / \chi$, $\chi$ is the comoving distance to redshift $z$ and $W^{i}_{\kappa}(z)$ is the lensing efficiency which is given by:
\begin{equation}
    W^{i}_{\kappa}(z) = \frac{3 H_0^2 \Omega_m}{2 c^2} \frac{\chi}{a(\chi)} \int_{\chi}^{\infty} d\chi' n^{i}_{\kappa}(z(\chi')) \frac{dz}{d\chi'}\frac{\chi' - \chi}{\chi'}.
\end{equation}
Here $n^{i}_{\kappa}$ is the normalized redshift distribution of the source galaxies corresponding to the tomographic bin $i$ (see \cite{paper1}). 

In order to model the matter multipole-space kernel we use the modeling framework similar to the one described in \citet{Mead:2015yca}, 
which is written as:
\begin{equation}\label{eq:halopowernu}
u_{\rm{m}}(k,M)=\sqrt{[1-\epow{-(k/k_*)^2}]} \ \frac{1}{\bar\rho}\ M \ W(\nu^{\eta_{\rm hm}} k,M),
\end{equation}
where, $\nu = \delta_{\rm sc}/\sigma(M)$ is the peak height, $\delta_{\rm sc}$ is the collapse threshold calculated from linear-theory and $\sigma(M)$ is the  standard-deviation of the linear density field filtered on scale containing mass $M$.  
The exponential factor inside the square root, depending on $k_*$, damps the one-halo term to prevent one-halo power from rising above linear at the largest scales (c.f., \citet{Mead_2021}). The parameter $\eta_{\rm hm}$ bloats the halo profiles, and we describe $W(k,M)$ below.

The halo window function, $W(k,M)$, has an analytical form for an NFW profile depending upon the halo concentration $c$ \citep{Cooray:2002}:
\begin{multline}
 W(k,M)\psi(c)=[\Ci(k_\mathrm{s}(1+c))-\Ci(k_\mathrm{s})]\cos(k_\mathrm{s}) \\
+[\Si(k_\mathrm{s}(1+c))-\Si(k_\mathrm{s})]\sin(k_\mathrm{s})-\frac{\sin(ck_\mathrm{s})}{k_\mathrm{s}(1+c)}\ ,   
\end{multline}
where $\psi(c)=\ln(1+c)-c/(1+c)$, $\Si(x)$ and $\Ci(x)$ are the sine and cosine integrals, $k_\mathrm{s}=kr_\mathrm{v}/c$ and $r_\mathrm{v}$ is the halo virial radius. The halo concentration is calculated by following the prescriptions of \citet{Bullock_2001} using:
\begin{equation}\label{eq:concentration}
c(M,z)=A_{\rm hm}\frac{1+z_\mathrm{f}}{1+z}\ , 
\end{equation} 
where $A_{\rm hm}$ is a free parameter. The formation redshift, $z_{\rm f}$, is then calculated using via \citep{press_74}:
\begin{equation}
\frac{g(z_\mathrm{f})}{g(z)}\sigma(\zeta M,z)=\delta_c\ ,
\label{eq:zf}
\end{equation}
where we fix $\zeta=0.01$ \citep{Bullock_2001, Mead:2015yca} 
and $g(z)$ is the growth function. We numerically invert the equation~(\ref{eq:zf}) to find $z_\mathrm{f}$ for a fixed $M$. Following the prescription of \citet{Mead:2015yca}, if $z_\mathrm{f}<z$, then we set $c=A_{\rm hm}$.

For the two-halo term, 
\begin{equation}\label{eq:bkl}
b_{\kappa}^{i}(\ell,z)=\frac{W^{i}_{\kappa}(z)}{\chi^2}  \ \sqrt{\left[1-f\tanh^2{(k\sigma_\mathrm{v}/\sqrt{f})}\right]},
\end{equation}
where $k = (\ell + 1/2) / \chi$ and we fix $f=0.188\times\sigma^{4.29}_8(z)$ \citep{Mead:2015yca}. The parameter $\sigma_\mathrm{v}$ denoting the 1D displacement standard deviation of the matter particles in linear theory is calculated via
\begin{equation}
\sigma^2_\mathrm{v}=\frac{1}{3}\int_0^{\infty}\frac{P_\mathrm{lin}(k)}{2\pi^2}\;\mathrm{d}\emph{k}\ .
\end{equation}

\subsection{Intrinsic Alignment Model}\label{sec:hm_IA}
The gravitational interaction of galaxies with the underlying dark matter field leads to their coherent alignment, also known as intrinsic alignments (IA) (see \cite{Troxel:2015} for a recent review). 
Since the alignments of galaxy shapes can be related to the underlying tidal field, intrinsic alignments can be described using perturbation theory \citep{Hirata:2004, Bridle:2007} or halo model \citep{Schneider:2010, FortunaIA:2021} frameworks. However, the detailed mechanism of IA depends on galaxy samples, their redshifts, host halo masses, and environments. The detailed modeling of IA, especially in one-halo and one-to-two halo transition regime, is an area of active study using data and simulations \citep{Hilbert_2017,Georgiou_2019,Samuroff:2019,samuroff2020advances,Okumura_2009,Mandelbaum_2006,Joachimi_2011,Johnston_2019,Singh_2015}. In this study, we model the effects of IA on our observable using the well studied non-linear alignment model (NLA) \citep{Bridle:2007}. This model is an effective two-halo model of IA and can be used to model the one-to-two halo transition scale and larger scales. We determine the scales over which this model is robust by comparing it to a halo model of IA as described below. We expect the halo model to be a better description of the small-scale intrinsic alignments, but it is computationally intensive to evaluate, and the specific analysis choices await future studies. Therefore, we determine the scales over which the NLA model of IA is a good approximation using the procedure described below.

In the halo model framework, the multipole space profile of intrinsic alignment is modeled as:
\begin{equation}
    \bar{u}^{i}_{\textrm{I}}(\ell,M,z) = f_s(z) \frac{n^{i}_{\kappa}}{\chi^2}\frac{dz}{d\chi} \frac{N_s(z,M)}{\bar{n}_s(z)} |\gamma^{I}_s(k,z,M)|,
\end{equation}
where $f_s(z)$ is the satellite fraction, $N_s(z,M)$ is the number of satellite galaxies in halo of mass $M$ at redshift $z$, $\bar{n}_s(z) = \int dM \frac{dn}{dM} N_s(z,M)$ is the number density of the satellite galaxies, and $|\gamma^{\rm I}_s(k,z,M)|$ is the density weighted ellipticity of the satellite galaxies.  We assume that we are dominated by blue galaxies in our source galaxy sample \citep{Samuroff:2019} and we model the satellite fraction, $f_s(z)$ as (see Fig.~A1 of \citet{FortunaIA:2021}):
\beqa
f_s(z) =
\begin{cases}
0.25 - 0.2z \,, \ \ \ z < 1.0 \\
0.05 \,, \ \ \   z > 1.0 
\, \nonumber
\end{cases}
\label{eq:fsz}
\eeqa

We model the number of satellite galaxies as: 
\begin{equation}
    N_s(z,M) = \frac{1}{2} \bigg[1 + {\rm erf}\bigg(\frac{ \log{M} - \log{M_{\rm min}}}{\sigma_{\log{M}}} \bigg) \bigg] \times \bigg( \frac{M_{\rm h}}{M_1} \bigg)^{\alpha_g}
\end{equation}
where we fix $\log{M_{\rm min}} = 11.57$, $\sigma_{\log{M}} = 0.17$, $\log M_1 = 12.75$ and $\alpha_g = 0.99$. For modeling $|\gamma^{I}_s(k,z,M)|$, we use Eq.16 of \citet{FortunaIA:2021}. However, in order to be conservative compared to the results of \citet{FortunaIA:2021} (to account for differences between the DES galaxies and their galaxy samples and modeling uncertainties), we use a large value of the amplitude of one-halo IA term $a_{1h} = 0.003$. 

The effective bias for the two-halo term is given by:
\begin{equation}
    b_{\textrm{I}}^{i}(\ell,z) = A(z) \,\frac{n^{i}_{\kappa}}{\chi^2} \frac{dz}{d\chi},
\end{equation}
where the IA amplitude is modeled using a power-law scaling as:
\begin{equation}\label{eq:AzIA}
    A(z) = -A_{\rm IA} \bigg(\frac{1+z}{1+z_0}\bigg)^{\eta_{\rm IA}} \frac{C_1 \rho_{\rm m,0}}{D(z)},
\end{equation}
and we set $z_0 = 0.62$ and $C_1 = 5\times 10^{-14} M_{\odot}^{-1}h^{-2}{\rm Mpc}^3$ \citep{Brown:2002}.

We model the one-halo correlations between Compton-$y$ and IA similar to Eq.~\ref{eq:Cl1h} with  $A = \textrm{I}$ and $B=y$. The two-halo term is modeled similar to Eq.~\ref{eq:Cl2h}, but in order to describe the correlations on smaller non-linear scales, we use the non-linear matter power-spectrum ($P_{\rm{NL}}(k,z)$) obtained from the \textsc{halofit} fitting function. This model is hence similar to the non-linear alignment model (NLA) as used previously in the calculation of the lensing cross-correlations:
\begin{equation}\label{eq:Cl_NLA}
C^{i j}_{\textrm{I} y;\textrm{NLA}}(\ell) = \int_{z_{\rm{min}}}^{z_{\rm{max}}} dz \frac{dV}{dz d\Omega} b_{\textrm{I}}^{i}(\ell,z) \ b^j_{y}(\ell,z) \ P_{\rm{NL}}(k,z).
\end{equation}

In order to mitigate systematic biases originating from complex inter-halo dynamics that might violate our assumptions described above, we use NLA as our fiducial intrinsic alignment model. We determine the scales that can be well described with this model through simulated analysis as described in \S\ref{sec:settings}. We compare theory $\xigty$ datavectors with no IA contributions, full halo model IA, $\xi_{\gamma_t y; \mathrm{HM}}^{ij}$, and NLA model IA, $\xi_{\gamma_t y; \mathrm{NLA}}^{ij}$ (see \S\ref{sec:IA_impact} for details).  
Note that in order to model halo exclusion and avoid double counting of non-linear information, when predicting $\xi_{\gamma_t y; \mathrm{HM}}^{ij}$ we truncate the two-halo contribution with a window function $f^{\rm 2h-trunc} = \exp{\big[ -(k/k_{\rm 2h})^2 \big]}$, where $k_{\rm 2h} = 6h/{\rm Mpc}$ \citep{FortunaIA:2021}.

\subsection{Final model for the shear-\textit{y} correlations}\label{sec:2pt_model_final}

The total model for the lensing-$y$ correlation is given by Eq.~\ref{eq:xigty}, where $C^{i}_{\kappa y;\textrm{model}}(\ell)$ is given by:
\begin{equation}\label{eq:fid_model}
    C^{ij}_{\kappa y;\textrm{model}}(\ell) = C^{i}_{\kappa y;\textrm{1h}}(\ell) + C^{ij}_{\kappa y;\textrm{2h}}(\ell) + C^{ij}_{\textrm{I} y;\textrm{NLA}}(\ell)
\end{equation}
We model the photometric uncertainity in our source redshift distribution $n^i_{\kappa}(z)$ using the shift parameters ($\Delta z^i_{\kappa}$) which modify the source redshift distributions as \citep{Krause_2017}:
\begin{equation}\label{eq:Delzi}
    n^i_{\kappa}(z) \rightarrow n^i_{\kappa}(z - \Delta z^i_{\kappa})
\end{equation}
We model the multiplicative shear calibration using:
\begin{equation}\label{eq:mi}
    \xi^{ij}_{\gamma_t y}(\theta) \rightarrow (1 + m^i) \, \xi^{ij}_{\gamma_t y}(\theta)
\end{equation}
We treat the four shift parameters $\Delta z^i_{\kappa}$ and four $m^i$ as free parameters and marginalize over them with Gaussian priors (see Table~\ref{tab:params_all}). 

In Fig.~\ref{fig:gty_sens} we show the sensitivity of the measured correlations to halo mass and redshift. We use the \textit{break model} to model the pressure profile and the parameter values of the full model (along with reference equations) are detailed in Table \ref{tab:params_all}. 
We plot results for several $\theta$ values. 
Due to the 10 arcmin smoothing applied to the $\planck$ $y$-map, cross-correlations between this map and $\des$ are dominated by contribution from halos with $M_{200c} > 10^{14} M_{\odot}/h$.  The significantly smaller beam of the ACT $y$-map (roughly 1.6 arcmin) means that cross-correlations between the ACT $y$-map and DES probe much lower halo masses and higher redshifts.

\begin{figure*}[ht]
\includegraphics[width=1.0\textwidth]{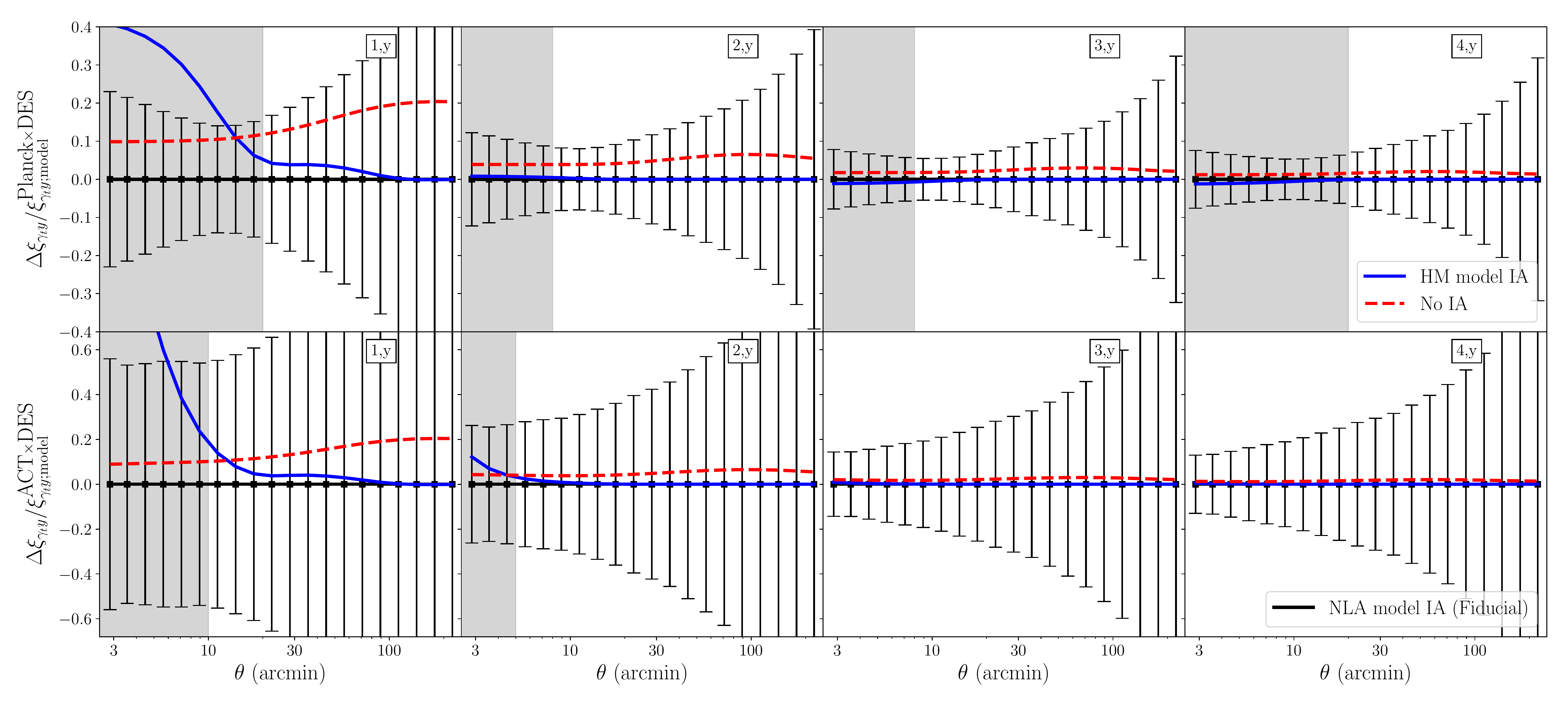}
\caption[]{Differences between the predicted shear-$y$ correlation ($\xi_{\gamma_t y}$) for different models of intrinsic alignment (IA), see \S~\ref{sec:hm_IA} for details.  
The quantity $\Delta \xi_{\gamma_t y}$ is the difference relative to our fiducial model (NLA), and we normalize all curves by this model. Note that due to the different beam sizes of the $\planck$ (top row) and ACT (bottom row) $y$-maps, the models for these two datasets are different.
The error-bars indicate the uncertainty on the model using the angular binning applied in the data analysis.  We see that in some cases, the difference between the models that include IA and the model without IA can approach a significant fraction of the uncertainty on the measurements. The gray regions indicate the scale cuts used in our analysis (see \S~\ref{sec:settings} for details). While determining these scale cuts, we impose the criteria that the difference in $\chi^2$ between the predictions from the two IA models is less than 1/8 (where $\chi^2$ is computed using the covariance used to analyze the data).  This ensures that the total difference in $\chi^2$ across all bins is less than one.
We restrict our analysis to scales larger than this threshold to minimize the impact of uncertainty in the IA model on our analysis.
}
\label{fig:ia_sc}
\end{figure*}

\begin{table}
\centering 
% \resizebox{\textwidth}{!}
\tabcolsep=0.16cm
\begin{tabular}{|c| c c c|}
\hline
% \hline
Model & Parameter & Fiducial, Prior & Equation  \\ \hline
% & & & \\
\multirow{16}{*}{\shortstack[c]{Common\\ Parameters}} & \multicolumn{3}{c|}{\textbf{Intrinsic Alignment}} \\ 
& $A_{\rm IA}$ & 0.5,$\mathcal{U}[-0.3,1.5]$   &  Eq.~\ref{eq:AzIA}    \\
& $\eta_{\rm IA}$ & 0.0,$\mathcal{U}[-3.0,4.0]$   &  Eq.~\ref{eq:AzIA}   \\ 
% & & & \\
\cline{2-4}
% & & & \\
& \multicolumn{3}{c|}{\textbf{Dark Matter Profile}} \\  
& $A_{\rm hm}$ & 2.32,$\mathcal{U}[0.1,5.0]$   & Eq.~\ref{eq:concentration}    \\
& $\eta_{\rm hm}$ & 0.76,$\mathcal{U}[0.1,1.0]$   &  Eq.~\ref{eq:halopowernu}  \\ 
% & & & \\
\cline{2-4}
% & & & \\
& \multicolumn{3}{c|}{\textbf{Shear Calibration}} \\ 
& \shortstack[c]{$m^{1}$}   & 0.0,$\mathcal{G}[-0.0063, 0.0091]$ & Eq.~\ref{eq:mi} \\ 
& \shortstack[c]{$m^{2}$}   & 0.0,$\mathcal{G}[-0.0198, 0.0078]$ & Eq.~\ref{eq:mi} \\ 
& \shortstack[c]{$m^{3}$}   & 0.0,$\mathcal{G}[-0.0241, 0.0076]$ & Eq.~\ref{eq:mi} \\ 
& \shortstack[c]{$m^{4}$}   & 0.0,$\mathcal{G}[-0.0369, 0.0076]$ & Eq.~\ref{eq:mi} \\ 
% & & & \\
\cline{2-4}
% & & & \\
& \multicolumn{3}{c|}{\textbf{Source photo-$z$}} \\  
& $\Delta z_{\kappa}^{1}$ & 0.0,$\mathcal{G}[0.0, 0.018]$ & Eq.~\ref{eq:Delzi}  \\ 
& $\Delta z_{\kappa}^{2}$ & 0.0,$\mathcal{G}[0.0, 0.015]$ & Eq.~\ref{eq:Delzi}  \\ 
& $\Delta z_{\kappa}^{3}$ & 0.0,$\mathcal{G}[0.0, 0.011]$ & Eq.~\ref{eq:Delzi}  \\ 
& $\Delta z_{\kappa}^{4}$ & 0.0,$\mathcal{G}[0.0, 0.017]$ & Eq.~\ref{eq:Delzi} \\ 
% & & & \\
\hline 
% & & & \\
% & & & \\
& \multicolumn{3}{c|}{\textbf{Pressure Profile}} \\  
\multirow{4}{*}{\shortstack[c]{\textit{Break Model}\\ $P_e \equiv P^{\rm B12, break}_e$}} & $P_0$ & 18.1,$\mathcal{U}[2.0,40.0]$ & Eq.~\ref{eq:B12_scaling} \\
& $\beta$ & 4.35,$\mathcal{U}[2.0,8.0]$ & Eq.~\ref{eq:B12_scaling}  \\
& $\alpha_z$ & 0.758,$\mathcal{U}[-6.0,6.0]$ & Eq.~\ref{eq:B12_scaling} \\  
& $\alpha^{\rm break}_{\rm m}$ & 0.0,$\mathcal{U}[-2.0,2.0]$   & Eq.~\ref{eq:Pe_total} \\
\hline
% & & & \\
& \multicolumn{3}{c|}{\textbf{Mass Bias}} \\  
\multirow{1}{*}{\shortstack[c]{\textit{Arnaud10} \\ \textit{Model 1}\\ $P_e \equiv P^{\rm A10c}_e$}} & $B$ & 1.4,$\mathcal{U}[0.9,2.8]$ & Eq.~\ref{eq:A10_model}  \\
& & & \\
& & & \\
\hline
% & & & \\
& \multicolumn{3}{c|}{\textbf{Mass Bias Redshift Evolution}} \\  
\multirow{2}{*}{\shortstack[c]{\textit{Arnaud10} \\ \textit{Model 2}\\ $P_e \equiv P^{\rm A10z}_e$}} & $B$ & 1.4,$\mathcal{U}[0.9,2.8]$ & Eq.~\ref{eq:Bz}  \\ 
& $\rho_{B}$ & 0.0,$\mathcal{U}[-3.0,3.0]$ & Eq.~\ref{eq:Bz}  \\ 
& & & \\
\hline
\end{tabular}
\caption{The parameters varied in different models, their prior range used ($\mathcal{U}[X, Y] \equiv$ Uniform prior between $X$ and $Y$; $\mathcal{G}[\mu, \sigma] \equiv$ Gaussian prior with mean $\mu$ and standard-deviation $\sigma$) in this analysis and the equations in the text where the parameter is primarily used.}
\label{tab:params_all}
\end{table}

\subsection{Covariance model}\label{sec:covariance}

We measure the cross-correlations of the DES shears with the $\act$ $y$-map and the $\planck$ $y$-map.  We leave a buffer region of approximately 6 degrees between the two $y$-maps to minimize covariance between the two measurements and ignore covariance between these two measurements below.  However, we do need to model the covariance between different angular and redshift bins.  

We model the covariance, $\varmathbb{C}$, of the shear and Compton-$y$ cross-spectra as a sum of Gaussian ($\varmathbb{C}^{\rm G}$) and non-Gaussian ($\varmathbb{C}^{\rm NG}$) terms.   The multipole-space Gaussian covariance is given by \citep{Hu:2004}: 
\begin{multline}\label{eq:cov_g}
\varmathbb{C}^{\rm G} (C^{i,j}_{\kappa,y}({\ell_1}),C^{l,j}_{\kappa,y}({\ell_2})) = \frac{\delta_{\ell_1 \ell_2}}{f^{(j)}_{\rm sky} (2 \ell_1 + 1)\Delta \ell_1} \\ \bigg[ \hat{C}^{il}_{\kappa \kappa}(\ell_1) \hat{C}^{jj}_{yy}(\ell_2) + \hat{C}^{ij}_{\kappa y}(\ell_1) \hat{C}^{lj}_{\kappa y}(\ell_2)  \bigg].
\end{multline}
Here, $\delta_{\ell_1 \ell_2}$ is the Kronecker delta, $f^{(1)}_{\rm sky} = 0.083$ for $\planck \times \des$ and $f^{(2)}_{\rm sky} = 0.0095$ for $\act \times \des$ are the effective sky coverage fractions; $\Delta \ell_1$ is the size of the multipole bin, and $\hat{C}_{\ell}$ is the total cross-spectrum between any pair of fields including the noise contribution: $\hat{C}_{\ell} = C_{\ell} + N_{\ell}$, where $N_{\ell}$ is the noise power spectrum of the field.  For the lensing convergence, we assume
\begin{equation}
N^i_{\kappa \kappa}(\ell) = \frac{\sigma^{2}_{e,i}}{{n^{i}_{\textrm{eff}}}},
\end{equation}
where 
$\sigma^{2}_{e,i}$ is the ellipticity dispersion and $n^{i}_{\textrm{eff}}$ is the effective number density of source galaxies, both in the $i$th source galaxy bin. For the $y$ field, we replace $\hat{C}_{yy}$ with the measured Compton-$y$ auto-power spectrum, which captures all the contributions from astrophysical and systematic sources of noise. We use the \texttt{NaMaster} \citep{Alonso_2019} algorithm to estimate this auto-power spectrum of both $\planck$ and $\act$ Compton-$y$ maps after accounting for their respective masks.

The non-Gaussian part can be written as
\beq\label{eq:cov_ng}
\varmathbb{C}^{\rm NG} (C^{i,j}_{\kappa,y}({\ell_1}),C^{l,j}_{\kappa,y}({\ell_2})) = \frac{1}{4 \pi f^{(j)}_{\rm sky}} \varmathbb{T}^{\rm i,j ; l,j}_{\kappa y; \kappa y}({\ell_1 \ell_2}),
\eeq
where we model only the 1-halo part of the trispectrum $\varmathbb{T}$ as that is expected to be dominant for the large halo masses that we are sensitive to \citep{Cooray_2001}. This term is modeled as:
\beq\label{eq:cov_trispec}
\varmathbb{T}^{\rm i,j ; l,j}_{\kappa y; \kappa y}({\ell_1 \ell_2}) = \int dz \frac{dV}{dz d\Omega} dM \frac{dn}{dM} \bar{u}_{\kappa}^{i}(\ell_1) \bar{u}_{y}^{j}(\ell_1) \bar{u}_{\kappa}^{l}(\ell_2) \bar{u}_{y}^{j}(\ell_2).
\eeq

Finally, we convert the multipole-space estimates of covariance to angular space using:
\begin{multline}\label{eq:cov_tot}
    \varmathbb{C}(\xi^{ij}_{\gamma_t y}(\theta_1),\xi^{lj}_{\gamma_t y}(\theta_2)) = \\
    \frac{1}{4\pi^2} \int \frac{d\ell_1}{\ell_1} \int \frac{d\ell_2}{\ell_2} \ell^2_1 \ell^2_2 J_2(\ell_1 \theta_1) J_2(\ell_2 \theta_2) \\
    \times \bigg[ \varmathbb{C}^{\rm G} (C^{i,j}_{\kappa,y}({\ell_1}),C^{l,j}_{\kappa,y}({\ell_2})) + \varmathbb{C}^{\rm NG} (C^{i,j}_{\kappa,y}({\ell_1}),C^{l,j}_{\kappa,y}({\ell_2})) \bigg]
\end{multline}
To evaluate these integrals, we use the fast-Fourier transform technique as detailed in \citet{Fang_2020}. We estimate our fiducial covariance matrix at $\planck$ cosmology and fiducial parameter values as described in Table \ref{tab:params_all}. The correlation matrix corresponding to our fiducial covariance is presented in Appendix~\ref{app:cov}. We refer the reader to \paperA $\,$ for details on validation of the covariance using simulations and jackknife procedure (this validated covariance is used in the data analysis of both papers). 

As described in \citet{osato2020supersample} using the Compton-$y$ auto-power spectrum, the trispectrum term (see Eq.~\ref{eq:cov_ng}, also referred to as connected non-Gaussian term, cNG) is the dominant contributor to the non-Gaussian covariance in Compton-$y$ correlations. The super-sample covariance makes a subdominant contribution in the presence of cNG due to large Poisson number fluctuations of massive clusters, and hence we ignore its contribution in this analysis (see \citet{osato2020supersample} for details).

\section{Data analysis}\label{sec:settings}

We do not expect our model to capture all physical effects over all angular scales.  For instance, we expect our fiducial intrinsic alignment model to break down at small scales due to complex non-linear processes impacting the tidal field and alignment of satellite galaxies. Even though we can remove the mean CIB contamination in our Compton-$y$ map using our constrained NILC methodology described in \S\ref{sec:measurement}, we expect other complex small-scale systematics like the variations in the CIB spectral energy distribution (SED) across the sky to contaminate our estimated $y$-maps.  
We prevent these effects from biasing our results by excluding those angular scales that are most impacted.  

\subsection{Impact of intrinsic alignments}
\label{sec:IA_impact}
A comparison of our shear-$y$ models with the halo model of IA ($\xi_{\gamma_t y; \mathrm{HM}}$), our fiducial NLA model  ($\xi_{\gamma_t y; \mathrm{NLA}}$), and without any IA contribution is shown in Fig.~\ref{fig:ia_sc}.  We also show the estimated error bars for $\planck \times \des$ and $\act \times \des$ in the figure, demonstrating our sensitivity to the IA model.  Especially for the first two tomographic bins, we see that the impact of IA can be significant relative to our error bars. Note that we use the value of $A_{\rm IA} = 0.5$ for the NLA model which is the mean of marginalized constraints obtained from DES-Y1 joint analysis of galaxy clustering and weak lensing \citep{des_y1_3x2pt}. Apparently, shear-$y$ correlations have now reached the sensitivity where the impact of IA should be included for an unbiased analysis; previous analyses of the shear-$y$ correlation have ignored the impact of IA. 

In order to mitigate the biases originating from the high-order intrinsic alignment process, we estimate the scales where our fiducial NLA model is a good approximation to a more complex halo model of IA (as described in \S\ref{sec:hm_IA}).
We use the halo model framework as described in \citet{FortunaIA:2021}, but we expect the specific parameter values of the model to be uncertain due to differences in the colors and environment of the source galaxies as well as due to the impact of baryonic physics, which was not modeled in their simulation-based study. Therefore, being conservative, we choose the values of the parameters describing the one-halo IA profile as three times the constraints in \citet{FortunaIA:2021}. The predicted theory curve with this configuration is shown using blue color in Fig.~\ref{fig:ia_sc}.

We restrict our fits to those angular scales for which the difference between our fiducial IA model and the halo-model model is small relative to our uncertainties.  In particular, we set a threshold total $\Delta \chi^2 = 1$ between NLA and halo-model simulated theory curves, and require that no single redshift bin contribute more than $1/N_{\rm bins}$ to the total $\Delta \chi^2$, where $N_{\rm bins}$ is the number of redshift bins in the analysis measured for both ACT and $\planck$ (i.e. $N_{\rm bins} = 8)$. 
For each tomographic cross-correlation $\xi_{\gamma_t y_j; \mathrm{NLA}}^{i}$, we find the minimum angular separation that satisfies our $\chi^2$ requirement
and exclude data points at smaller separations. In calculating this $\Delta \chi^2$ per bin, ${\varmathbb{C}}_{ij}$ is the covariance matrix corresponding to that specific tomographic bin and scales greater $\theta_{\textrm{sc}}^{ij}$. 

Note that the curve with zero-IA contribution in Fig.~\ref{fig:ia_sc} lies above the one with fiducial IA contribution. In simple galaxy alignment models, the galaxies are typically aligned in the stretching direction of the tidal field, while the gravitational shearing occurs in tangential direction that is traced by tSZ \citep{Hirata:2004,Troxel:2015}.
% The direction of IA is typically orthogonal to the tangential direction of gravitational shearing that tSZ traces \citep{Troxel:2015}. 
This leads to an anti-correlation between IA and tSZ that is followed by our fiducial model as well as our best fit model (see Fig.~\ref{fig:xigty_measure}). However, baryonic physics, galaxy infall and merger history can complicate this interpretation and can lead to a positive correlation. Therefore, we vary the coefficient of the IA model with a flat prior, allowing for both positive and negative values (see Table~\ref{tab:params_all}).

\subsection{Impact of CIB}

We also find that scales below 20~arcmin in the correlations between the last tomographic bin of DES shear catalog and $\planck$ $y$-map are impacted by the leakage of CIB. Additionally, we also remove the scales below 7~arcmin for all the tomographic bins of $\planck \times \des$, due to the impact of the non-trivial structure of the DES Y3 mask in the $\planck$ footprint on the small scales covariance between $\planck \times \des$ (see \paperA $\,$ for details on the impact of CIB and covariance validation). 
Note that, as the $\planck$ Compton-$y$ map has a beam of 10~arcmin, the smaller scales are heavily correlated, and we do not lose any appreciable signal-to-noise (see Fig.~\ref{fig:corr_mat}). After the scale cuts, we are left with $N_{\rm data} = 123$ points in our final datavector.

\subsection{Bayesian analysis}
\label{sec:bayesian_settings}
We perform our analysis at fixed cosmology, but explore the impact of using a different cosmological parameter choice on our results.  Our baseline analysis uses the best-fit flat $\Lambda$CDM model from \cite{Planck:2018cosmo}, with $\Omega_{\rm m} = 0.315$, $\sigma_8 = 0.811$, $H_0=67.4$, $\Omega_{\rm b}=0.0486$ and $n_{\rm s}=0.965$.  We test the impact of changing the cosmological parameters $\Omega_m$ and $\sigma_8$, which are the parameters Compton-$y$ correlations are most sensitive to \citep{Hill-Pajer2013,Komatsu_2002}. To that end we use $\des$ Year 1 constraints obtained from the joint analysis of galaxy clustering and lensing, $\Omega_{\rm m} = 0.264$ and $\sigma_8 = 0.807$ \citep{des_y1_3x2pt}. 

We list the set of parameters we vary in Table~\ref{tab:params_all} along with the priors used. We use wide uninformative uniform priors on all the parameters except shear calibration and source photo-$z$ shift parameters. We refer the reader to \citet{y3-sompz} and \citet{y3-imagesims} for details on the estimation of priors on the shear calibration and source photo-$z$ shift parameters. 

We assume the likelihood to be a multivariate Gaussian:
% \begin{linenomath*}
\begin{equation}
    \ln \mathcal{L}(\mathbfcal{D}|\Theta) = -\frac{1}{2} \left[\vec{\mathbfcal{D}} - \vec{\mathbfcal{T}}(\Theta) \right]^{\rm T} \, {\varmathbb{C}}^{-1} \,  \left[\vec{\mathbfcal{D}} - \vec{\mathbfcal{T}}(\Theta)\right].
\end{equation}
% \end{linenomath*}
Here $\vec{\mathbfcal{D}}$ is the measured $\xigty$ correlation datavector, with length $N_{\rm data}$, $\vec{\mathbfcal{T}}$ is the theoretical prediction for the cross-correlation at the parameter values given by  $\Theta$, and ${\varmathbb{C}}^{-1}$ is the inverse covariance matrix.% of shape $N_{\rm data} \times N_{\rm data}$.

We use \textsc{Polychord} \citep{Handley:2015:} to draw samples from the posterior:
\begin{equation}
    \mathcal{P}(\Theta | \mathbfcal{D}) \propto \mathcal{L}(\mathbfcal{D}|\Theta) {\rm P}(\Theta) %\frac{\mathcal{L}(\mathbfcal{D}|\Theta) {\rm P}(\Theta)}{{\rm P}(\mathbfcal{D})}
\end{equation}
where ${\rm P}(\Theta)$ are the priors on the parameters of our model.
We use 128 live-points as the settings of the \textsc{Polychord} sampler and set the length of the slice sampling chain to produce a new sample as 30. Convergence is declared when the total posterior mass inside the live points is 0.01 of the total calculated evidence. We note that the common parameters in Table~\ref{tab:params_all} and the likelihood sampler settings are same between \paperA $\,$ and this paper.

\section{Results}
\label{sec:results}

\begin{figure*}[ht]
\includegraphics[width=1.0\textwidth]{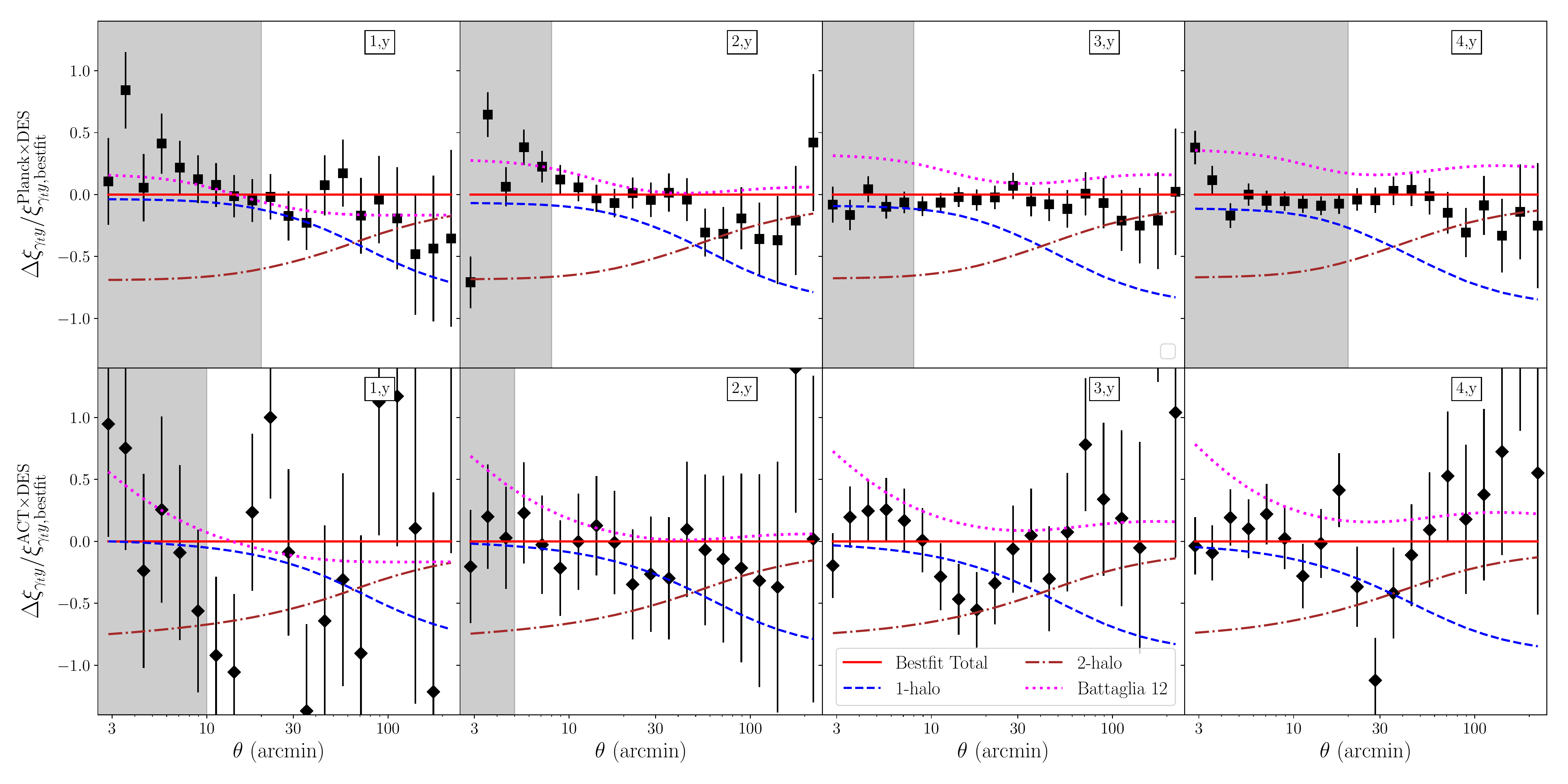}
\caption[]{Residuals of the best-fit to the $\planck \times \des$ (top) and $\act \times \des$ (bottom) shear-$y$ correlation measurements, using the \textit{break model} of pressure profile (see \S\ref{sec:model_y}).  Different columns represent the different redshift bins of the lensed source galaxy sample. We show the contributions to the total best-fit from 1-halo and 2-halo terms using blue dashed and brown dot-dashed curves (see Eq.~\ref{eq:fid_model}). We also compare with the predictions for shear-$y$ correlations when using preferred values of the pressure profile parameters from \citet{Battaglia:2012} fitting function with magenta dotted line.
}
\label{fig:residual_besfit}
\end{figure*}

\begin{figure}[ht]
\includegraphics[width=\columnwidth]{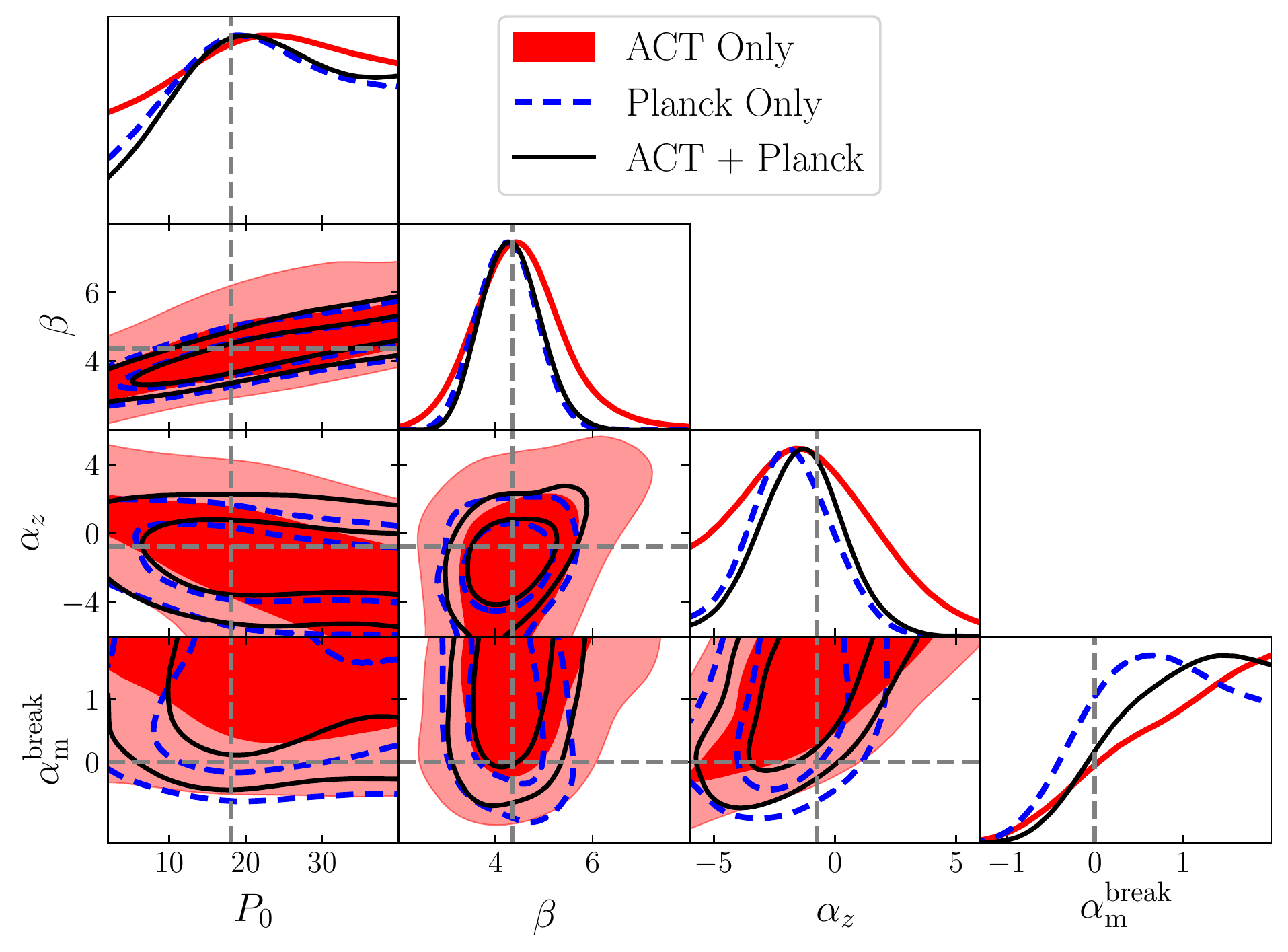}
\caption[]{Constraints on the pressure profile parameters from the \textit{break model} when using the Compton-$y$ map from $\act$ only, $\planck$ only and both. The gray dashed lines indicate the preferred values of the parameters from \citet{Battaglia:2012} fitting function.}
\label{fig:constraints_param}
\end{figure}

We now present the results of our analysis for the pressure profile models introduced in \S\ref{sec:model_y}: the \textit{break model} and the \citet{Arnaud:2010} model. We first analyze our measurements using the \textit{break model}, obtaining the parameter constraints of this generalized NFW model, inferring physical observables from these constraints and comparing them with previous studies. Lastly, we present the constraints on the hydrostatic mass bias parameter using the \citet{Arnaud:2010} model and compare with previous studies. 

\subsection{\textit{Break model}}
\subsubsection{Parameter constraints}

In Fig.~\ref{fig:residual_besfit} we show the residuals of our fit to the data using the \textit{break model} as described in \S\ref{sec:model_y}. We also show the one-halo and two-halo contributions to the total best-fit curve. Note that the contribution from the one-halo term extends out to large angular scales.
This behavior is because the lensing-$y$ correlation is sensitive to massive halos, and that $\gamma_t$ is a non-local quantity, with $\gamma_t$ at a scale $\theta$ sensitive to the correlation function at scales below $\theta$.
Also note that for the first two tomographic bins, the sum of the one-halo and two-halo contributions is more than the total best-fit curve; this is a consequence of intrinsic alignments in our best-fit model, which acts to suppress the correlation functions.

Our best fit yields a total $\chi^2 = 150.2$ with $N_{\rm data} = 123$ data points, which corresponds to a probability-to-exceed (PTE) of $0.033$ after accounting for the number of constrained model variables. In order to estimate the total constrained parameters, we compare the parameter constraints to the prior as described in \citep{Raveri_2019}.\footnote{We use the publicly available \texttt{tensiometer} code at \href{https://tensiometer.readthedocs.io/}{https://tensiometer.readthedocs.io/}} The somewhat high value of $\chi^2$ appears to be driven at least partly by the large-scale measurements of the shear-$y$ correlation with ACT.  Excluding scales above 100 arcmins for these measurements yields a PTE of $0.1$. As the \textsc{D56} region that the ACT Compton-$y$ map covers is near the galactic plane, there could be additional sources of noise that are not modeled in our fiducial covariance. We note that we have verified that our main conclusions in the following sub-sections are robust to this low PTE value, since they refer to low mass halos that are probed by small scales which are well fit with our models and also dominate the signal-to-noise. We also show the Arnaud et al. profile model (see  \S\ref{sec:model_y}) in Appendix.~\ref{app:fitA10} and find that to result in similar PTE values.

We also note that in the residuals shown in Fig.~\ref{fig:residual_besfit}, we see some evidence for departures from the model near the one-to-two halo transition regime. We find slight preference for higher pressure at the transition scales, which is particularly evident in top panels for $\planck \times$DES. 
Our model for the shear-$y$ correlation ignores the impact of shocks, which have recently been shown to impact the outskirts of stacked $y$ profiles of galaxy clusters \citep{Baxter:2021:}, and could therefore impact the shear-$y$ correlation measurements in the one-to-two halo regime. Additionally, the assumption used in this study that the linear halo bias model describes the 2-halo correlations can be broken near the transition regime due to non-linear effects of gravity. However, given that the PTE found in our fiducial analysis is not very low, we do not pursue these possibilities further and leave them to a future study.

In Fig.~\ref{fig:constraints_param} we show the constraints on the pressure profile parameters of the \textit{break model}. The full constraints for this model at both $\planck$ and $\des$-Y1 cosmologies on all the parameters (other than shear calibration and photo-z shift parameters, as they are prior dominated) are shown in Fig.~\ref{fig:cosmo_impact} in Appendix~\ref{app:cosmo_comp}. We find the constraints from analyzing the $\planck$-only and $\act$ correlations to be consistent.  The correlations with the $\planck$-only map have a higher total signal to noise owing to the larger area.  Note, though, from Fig.~\ref{fig:xigty_measure} that the smaller beam size of $\act$ equates to higher sensitivity to low mass and high-redshift halos.  

Our results exhibit a strong degeneracy between $P_0$ and $\beta$, making the marginalized posterior on $P_0$  very weak and the marginalized posterior on $\beta$ somewhat sensitive to our $P_0$ prior.  The redshift evolution parameter, $\alpha_z$, and the power-law index below the break mass, $\alpha_{\rm m}^{\rm break}$, are weakly constrained when using both the $\act$ and $\planck$ maps.  The dashed line in Fig.~\ref{fig:constraints_param} indicates the parameter values corresponding to the \citep{Battaglia:2012} model.

\subsubsection{Inferred redshift and mass dependence of the pressure profiles}

\begin{figure}[h]
 \includegraphics[width=\columnwidth]{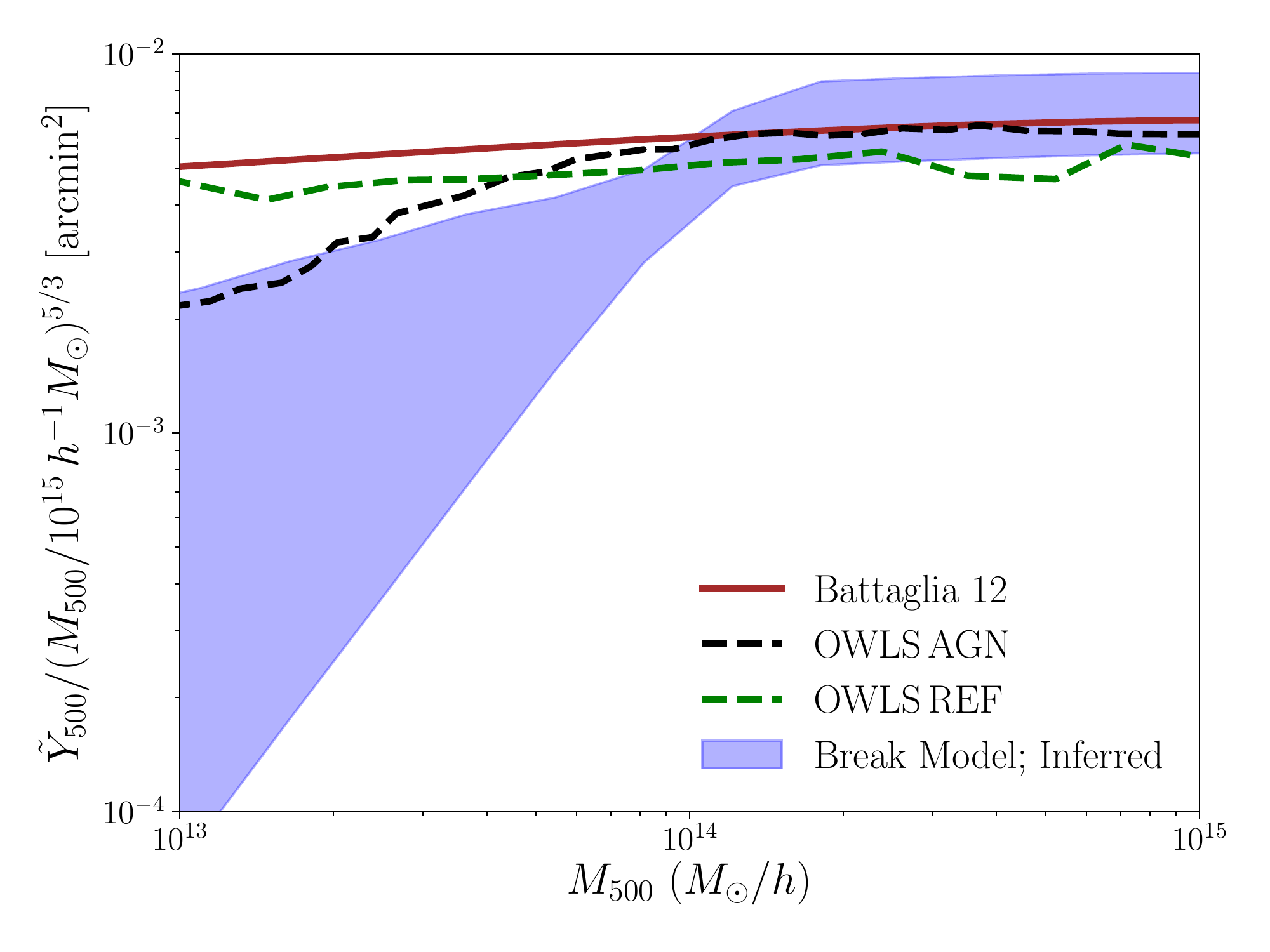}
\caption[]{Inferred 68\% credible interval (blue shaded region) on the $\tilde{Y}_{500} - M_{500}$ relation at $z=0.25$ using the \textit{break model}.  We compare predictions from various hydrodynamical simulations (curves). We find our inferences to be consistent with all the hydrodynamical simulations at high mass, but we find a departure for lower mass halos where AGN feedback has its greater impact.  
}
\label{fig:ym_comp}
\end{figure}

\begin{figure}[h]
\includegraphics[width=\columnwidth]{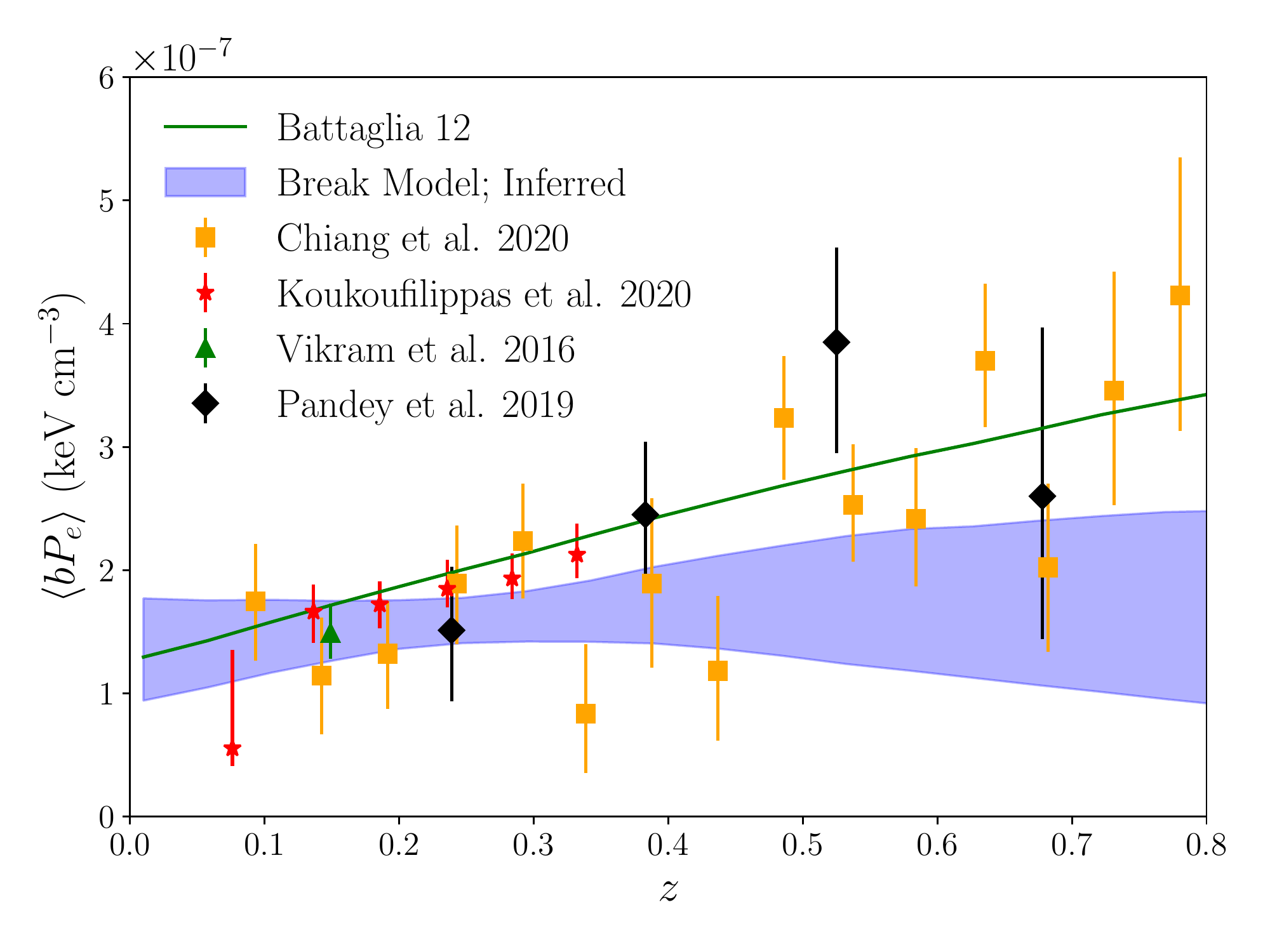}
\caption[]{Inferred 68\% credible interval (blue shaded region) on the bias weighted pressure of the universe ($\langle bP_e \rangle$) from our pressure profile constraints, assuming the \textit{break model}. We compare this inference to previous studies where constraints are obtained from cross-correlations between galaxy/halo catalogs with Compton-$y$ maps.}
\label{fig:bP_comp}
\end{figure}

\begin{figure}[h]
\includegraphics[width=\columnwidth]{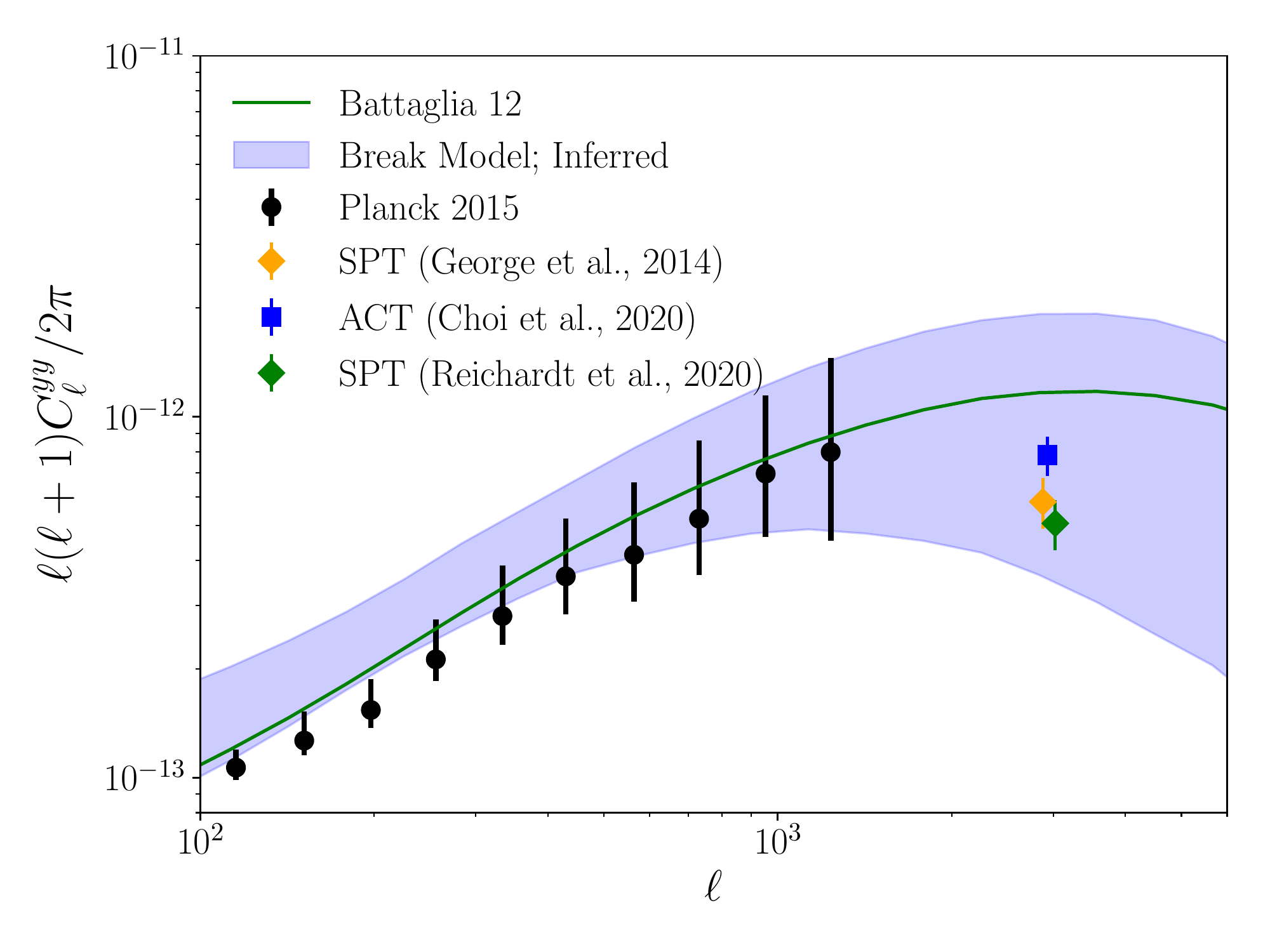}
\caption[]{Inferred 68\% credible interval (blue shaded region) on the auto-power spectra of Compton-$y$ when adopting the \textit{break model}. We compare this inference with measurements from the $\planck$, $\act$ and SPT Collaborations, finding a good agreement across all scales.  Our measurement is also consistent with expectations from the model of \citet{Battaglia:2012} (green curve).}
\label{fig:clyy_comp}
\end{figure}

We can translate the model posterior from our fits to the shear-$y$ correlation into constraints on the relation between the integrated halo $y$ signal and halo mass.  In Fig.~\ref{fig:ym_comp} we show the  $\tilde{Y}_{500} - M_{500}$ relationship inferred from the \textit{break model} fits, where $\tilde{Y}_{500}$ is given by:
\beq\label{eq:y500}
\tilde{Y}_{500}(M,z) = \frac{D^2_A(z) }{(500 {\rm Mpc})^2 E^{2/3}(z)} 
\frac{\sigma_T}{m_e c^2} \int_0^{R_{500c}} dr 4\pi r^2 \frac{P_e(r|M,z)}{D^2_A(z)},
\eeq
where $E(z)$ is the dimensionless Hubble parameter. In order to obtain the blue-shaded band in Fig.~\ref{fig:ym_comp}, we estimate the $\tilde{Y}_{500} - M_{500}$ relationship for 2000 samples from the posterior of the \textit{break model} and estimate the 68\% credible interval from the resulting curves. 

We compare the inferred $\tilde{Y}_{500} - M_{500}$ relationship from data to the predictions from various hydro-dynamical simulations incorporating different feedback mechanisms. The OWLS REF and OWLS AGN curves correspond to the cosmo-OverWhelmingly Large Simulation (cosmo-OWLS) simulations \citep{LeBrun:2014, McCarthy:2014}. OWLS REF includes the prescriptions for radiative cooling and supernovae feedback while OWLS AGN additionally includes the feedback from active AGN. The Battaglia 12 curve is derived from the \citet{Battaglia:2012} model.  This model also incorporates prescriptions for feedback mechanisms from supernovae and AGN feedback, but because it was calibrated at cluster-scale halo masses, we do not expect these effects to be captured correctly at low halo mass. We find that at higher masses, our inferred constraints agree with all three predicted pressure profile models. However, we find evidence for a decline in $\tilde{Y}_{500}$ for halos with mass $M < 10^{14} M_{\odot}/h$ compared to predictions from \citet{Battaglia:2012} and the OWLS REF simulations. We find that our constraints are in better agreement with OWLS AGN simulations. Note that \citet{Hill:2018} also found similar results using the cross-correlation of galaxies with $y$.

We also predict the evolution of the bias weighted average pressure of the universe ($\langle b P_e \rangle$) from our \textit{Break Model} constraints using:
\beqa\label{eq:bpe}
\langle b P_e\rangle (z) = (1+z)^3 \int_0^{\infty} \frac{dn}{dM} b(M,z) E_{\rm T}(M,z) dM,
\eeqa
where the total thermal energy of halo of mass $M$ at redshift $z$ is given by:
\begin{equation}
E_{\rm T}(M,z) = \int_0^{\infty} dr \, 4 \pi r^2 P_e(r,M,z).
\label{eq:press_avg}
\end{equation}

Here $P_e(r,M,z)$ are predicted using the samples from the posterior using Eq.~\ref{eq:UB}. The inferred constraints on $\langle b P_e \rangle$ following above methodology is shown in the blue band in Fig.~\ref{fig:bP_comp}. 
We compare our predictions to the previous studies that estimated $\langle b P_e \rangle$ by analyzing cross-correlations between Compton-$y$ and cluster catalogs \citep{Vikram2017} or galaxy catalogs \citep{Pandey:2019, chiang2020thermal, Koukoufilippas:2020} . We find a good agreement in our inference and previous studies at lower redshift with a mild deviation at higher redshift. Note that at higher redshifts ($z > 0.7$), $\langle b P_e \rangle$ receives a contribution from lower-mass halos (see Fig.~1 of \citep{Pandey:2019}) that our analysis is less sensitive to. We also note that our inference assumes the validity of the halo model to even small mass halos, and hence this methodology will miss the contribution in the filaments between large clusters. These caveats can qualitatively explain the mild deviation between our inference and previous measurements at high redshift.

Next, we propagate our parameter constraints to the auto-power spectra of Compton-$y$. The inferred constraints are shown using the blue band in Fig.~\ref{fig:clyy_comp}. We compare these predictions to the measurements from the Compton-$y$ maps from $\planck$ \citep{Planck:2015comptony} at larger scales. At smaller scales, we compare our inferences with estimates from $\act$ \citep{Choi:2020} and the South Pole Telescope (\textsc{SPT}) Collaboration \citep{Reichardt:2020} obtained from analyzing CMB data. We find that our inferences using the \textit{break model} is consistent with all the measurements. We also show the prediction from the \citet{Battaglia:2012} model.
While this simulation curve provides a good fit to the $\planck$ measurements, it over-predicts the auto-power spectrum at high multipoles that are dominated by high-redshift and low-mass halos. This figure highlights that inferences made using imminent higher significance measurements of the shear-$y$ cross-correlations, particularly in the small scales from ACT and SPT, will be crucial in establishing the consistency of the probe with Compton-$y$ auto-correlations and comparisons with simulations.

\begin{figure*}[ht]
\includegraphics[width=0.98\textwidth]{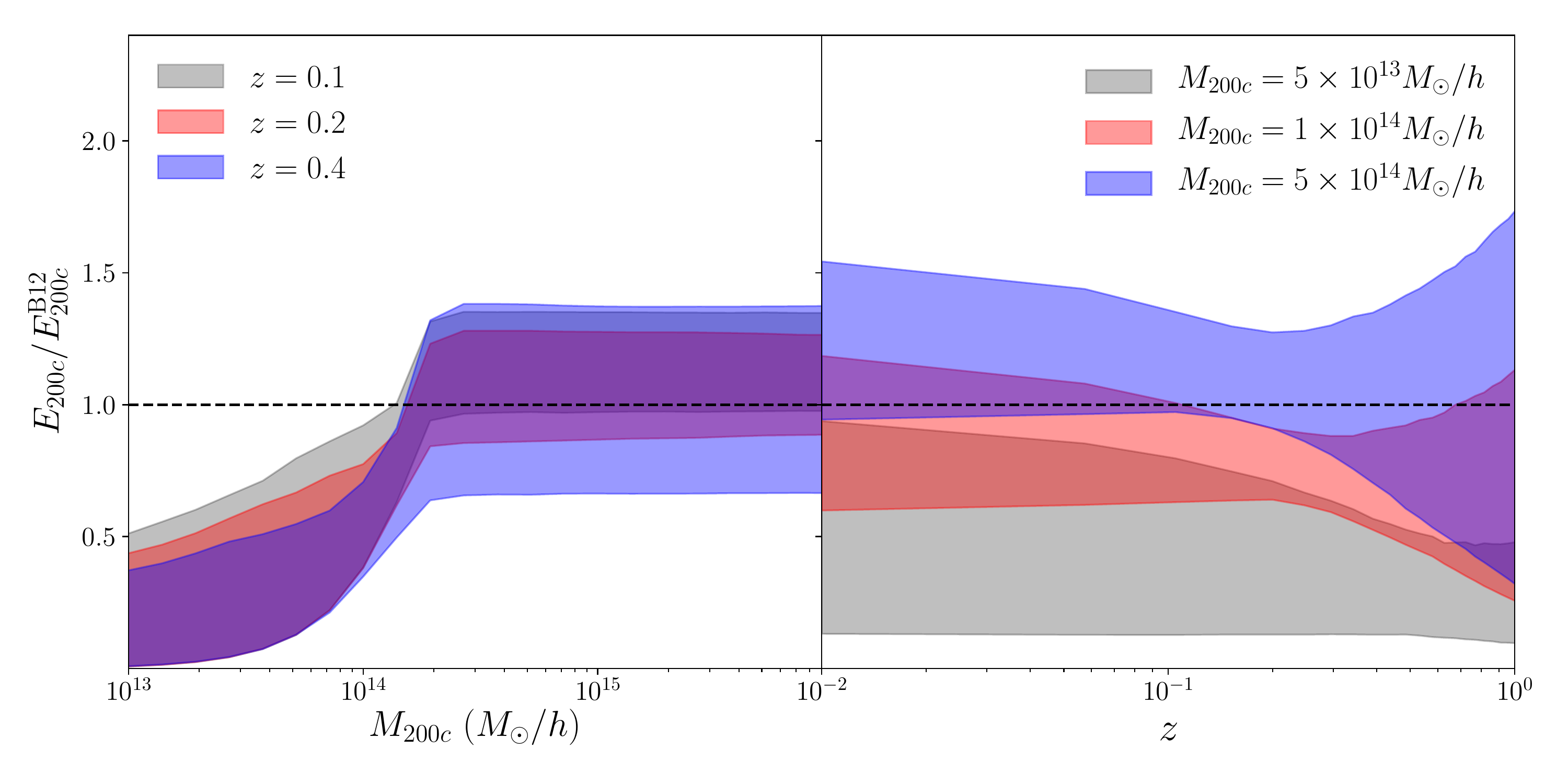}
\caption[]{
Constraints on the total thermal energy within $r_{200c}$ (see Eq.~\ref{eq:E200c}) of hot gas in halos inferred from the \textit{break model} analysis. We compare our constraints to the simulation-based predictions of \citet{Battaglia:2012}, finding good agreement at high halo mass but differences at low mass.  
}
\label{fig:pe_3d_comp}
\end{figure*}

We now use our inferred model constraints to generate constraints on the pressure profiles of halos as a function of mass and redshift.  
In Fig.~\ref{fig:pe_3d_comp} we show our constraints on the total thermal energy of hot gas inside $r_{200c}$:
\begin{equation}\label{eq:E200c}
E_{200c}(M,z) =  4\pi  \,\int_{0}^{r_{200c}} dr \, r^2 \, P_e(r,M,z),    
\end{equation}
with similar predictions using the \citet{Battaglia:2012} model (labeled $E^{\rm B12}_{200c}$). We find good agreement between our inferences and the simulation prediction for higher masses and lower redshift halos. However we see a clear departure from simulation predictions in lower mass halos. We find our inferences on the ratio $E_{200c}/E^{\rm B12}_{200c}$ are discrepant from unity in the mass range $10^{13} < M_{200c}(M_{\odot}/h) < 2\times 10^{14}$ at 3.0$\sigma$, 4.0$\sigma$ and 5.4$\sigma$ for $z = 0.1, 0.2$ and $0.4$ respectively (see the left panel of Fig.~\ref{fig:pe_3d_comp}). Similar conclusions were reached when extrapolating the tSZ analysis around Sloan Digital Sky Survey (SDSS) galaxy samples to smaller radii (see \citet{amodeo2020atacama, Schaan_2021}). However note that our sensitivity to the host halo masses and redshifts of the relevant SDSS galaxies used by \citet{amodeo2020atacama} is small. Moreover, they report excess pressure compared to the predictions from the \citet{Battaglia:2012} model outside of the virial radius of the halos. This behavior can occur due to ejection of hot gas from inside the halos due to feedback processes, which can lower the pressure inside the halos while raising it outside the virial radius.

\subsection{Mass bias constraints}

As described in \S~\ref{sec:model_y}, estimating the pressure profile of hot gas in halos gives a handle on its mass estimation. This is typically done using the \citet{Arnaud:2010} profile (see Eq.~\ref{eq:A10_model}), assuming the hot gas exists in hydro-static equilibrium. However, several physical processes (e.g., the flow of gases in filaments) can violate this assumption and bias the mass calibration. This bias is captured using a mass bias parameter $B$ and is typically studied in cluster mass scale halos. As the shear-$y$ cross-correlation is sensitive to these high-mass, cluster-scale halos (see Fig.~\ref{fig:gty_sens}), we can infer the hydro-static mass bias from our measurements and compare them with previous studies. Calibrating cluster masses is difficult, and some recent methodologies have lead to mild tension with the $\Lambda$CDM cosmology obtained from primary CMB power spectra analysis from the $\planck$ Collaboration \citep{PlanckCollaboration:2016,Zubeldia_2019, Costanzi_2019, DESy1_clusters, Bocquet_2019, Hasselfield_2013,Planck:2015comptony}. This uncertainty in cluster mass calibration is the leading systematic in obtaining cosmology from cluster counts (see e.g. \cite{McClintock_2018,Miyatake_2019,Murata_2018,von_der_Linden_2014,Hoekstra_2015}). The tSZ cross-correlation analysis studied here can provide an independent handle on this calibration.

\begin{figure}[h]
\includegraphics[width=\columnwidth]{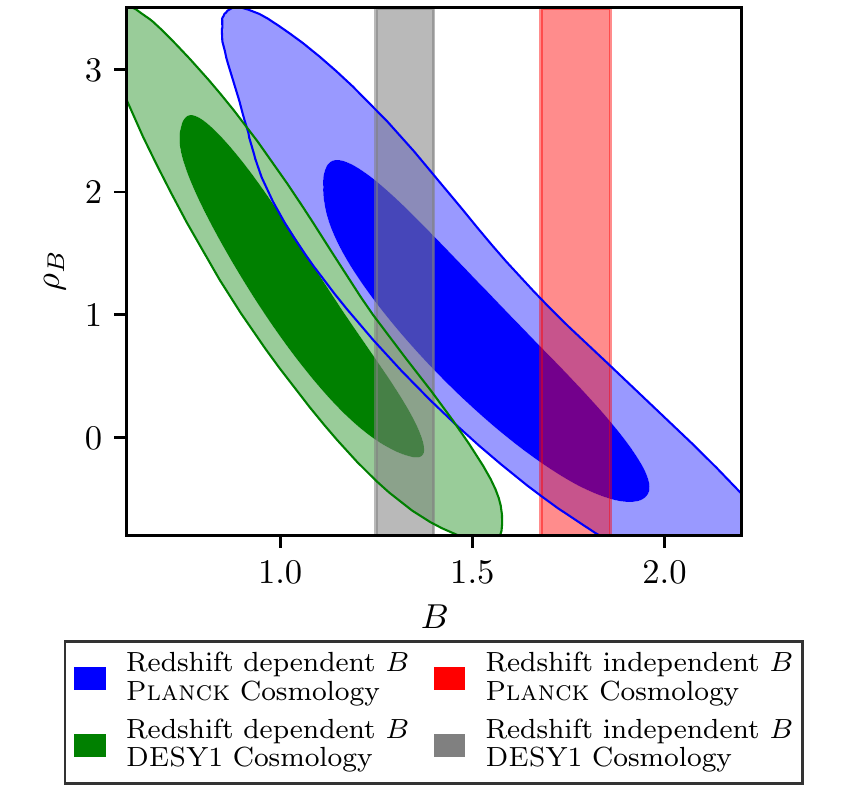}
\caption[]{Constraints on the mass bias and its redshift evolution using shear-$y$ cross-correlations. The red and gray vertical bands show the constraints on a constant mass bias parameter using the $P^{\rm A10c}_e$ model at the $\planck$-preferred and the DES-preferred cosmologies respectively. The blue and green contours corresponds to the $P^{\rm A10z}_e$ model (see Eq.\ref{eq:A10_model}) with mass bias evolving with redshift as $B(z) = B(1+z)^{\rho_B}$ at the $\planck$-preferred and the DES-preferred cosmologies respectively.}
\label{fig:hydroMB_comp}
\end{figure}

\begin{figure}[h]
\includegraphics[width=\columnwidth]{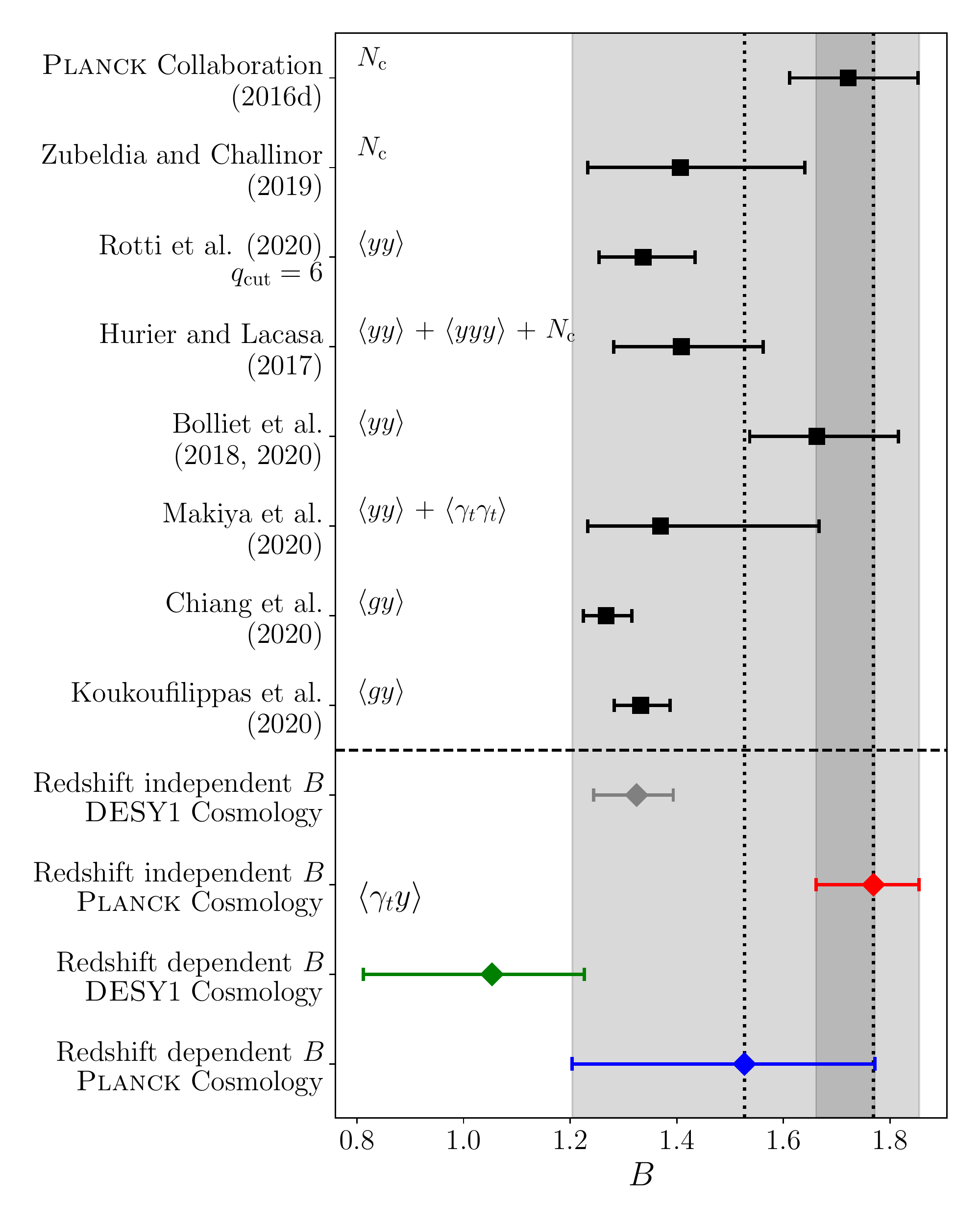}
\caption[]{Comparison of constraints on mass bias from this work and previous studies. The dashed black line and gray-bands correspond to the marginalized mean and uncertainty on the mass bias parameter obtained using both the redshift-independent and redshift-dependent mass bias models at the $\planck$-preferred cosmology.}
\label{fig:hydroMB_comp_lit}
\end{figure}

In Fig.~\ref{fig:hydroMB_comp}, at $\planck$ cosmology and with a model assuming a redshift independent mass bias parameter, we obtain marginalized constraints of $B = 1.8^{+0.1}_{-0.1}$, which translates to large $b_{\rm HSE} = (B - 1)/B = 0.4^{+0.03}_{-0.04}$. In Fig.~\ref{fig:hydroMB_comp_lit}, we compare our constraints obtained using shear-$y$ cross-correlations ($\langle \gamma_t y \rangle$) with previous studies based on the combinations of various observables, like cluster abundance ($N_c$), Compton-$y$ auto-power spectra ($\langle yy \rangle$), Compton-$y$ bispectra ($\langle yyy \rangle$), shear-2pt auto-correlations ($\gamma_t \gamma_t$) and cross-correlations between galaxy overdensity and Compton-$y$ ($\langle gy \rangle$). 

We find that our constraints on a redshift-independent mass-bias for the $\planck$ cosmology is consistent with previous analysis using tSZ cluster abundances and Compton-$y$ power spectra \citep{Planck:2015comptony,PlanckCollaboration:2016,Bolliet_2018,Bolliet_2020}. The cluster abundance and Compton-$y$ power spectra are largely sensitive to high mass halos which occupy lower redshifts. While we do expect a non-zero mass bias due to non-thermal pressure support of hot gas in halos, this mass bias value is large compared to the predictions from hydrodynamical simulations \citep{Biffi:2016} as well as analytical calculations \citep{Shi:2014} (typically preferring $b_{\rm HSE}  \in [0.1,0.2]$). Alternatively, this inconsistency can also be cast into the $\sigma_8$ parameter due to degeneracy between $B$ and $\sigma_8$.  Several low-redshift probes prefer a lower value of $\sigma_8$ compared to the constraints from primary CMB anisotropy analysis by $\planck$ \citep{Hikage_2019,Asgari_2021,des_y1_3x2pt}. Hence lowering the value of preferred $\sigma_8$ can result in a lower value of the mass bias parameter. A previous study by \citet{Zubeldia_2019} based on weak lensing based mass calibration, sensitive to lower redshifts, has reported a lower value of the mass bias as well as a lower value of $\sigma_8 = 0.76^{+0.04}_{-0.04}$ (see their paper for caveats about priors on Compton-$y$ scaling relations). Similarly other studies using weak lensing based mass calibration and richness-based mass calibrations have also reported a preference for lower mass bias \citep{von_der_Linden_2014, Hoekstra_2015,Miyatake_2019, Hilton_2018, Hilton_2021}. For example, in a recent analysis detailing updated ACT cluster catalog, \citet{Hilton_2021} estimated $b_{\rm HSE} = 0.31^{+0.07}_{-0.07}$ for clusters lying in the DES footprint with measured richness and using richness-mass relation as described in \citet{McClintock_2018}.\footnote{Note that this updated value of $b_{\rm HSE}$ is obtained from ACT DR5 catalog documentation detailed in \href{https://lambda.gsfc.nasa.gov/product/act/actpol_dr5_szcluster_catalog_info.cfm}{\text{https://lambda.gsfc.nasa.gov/product/act/actpol\_dr5\_szcluster\_catalog\_info.cfm}} and differs slightly from the value published in \citet{Hilton_2021}.} In a study by \citet{Hurier_2017}, jointly analyzing Compton-$y$ auto power spectra, bispectra and cluster abundances has also reported a lower value of mass bias and $\sigma_8 = 0.79^{+0.02}_{-0.02}$ which is still in mild tension with hydrodynamical and analytical estimates on $B$. In Fig.~\ref{fig:hydroMB_comp} we also find a lower value of redshift independent $B$ when using DES-Y1 cosmological parameters which prefers a lower value of $\sigma_8$ and $\Omega_{\rm m}$ (see \S\ref{sec:bayesian_settings}). This sensitivity of the mass bias parameter to cosmological parameters demands a study jointly constraining cosmological parameters and pressure profiles of halos. Note that the mass bias cannot be jointly constrained with cosmological parameters from our observable ($\langle \gamma_t y \rangle$) alone due to a large degeneracy between $\sigma_8$ and $B$. We defer the joint analysis of our observable with other observables, like shear-2pt auto-correlations to a future study.

As our source galaxy sample is divided into multiple redshift bins, we can probe the change in mass bias parameter with redshift using our tomographic datavector. While allowing for this redshift evolution, we obtain $B = 1.5^{+0.3}_{-0.3}$ at $z=0$, which translates to $b_{\rm HSE} = 0.34^{+0.1}_{-0.2}$ for the $\planck$ cosmology. With this model, the power-law index of the evolution of mass bias with redshift is found to be $\rho_B = 0.8^{+0.8}_{-1.0}$. As is shown in Fig.~\ref{fig:hydroMB_comp}, this model results is strong degeneracy between $B$ and $\rho_B$, hence degrading the error bars on $B$ significantly. However, the shift in the mean parameter values are such that this model makes the mass bias estimate at low redshift consistent with the estimates from previous studies using analytical calculation and simulations mentioned above as well as from cross-correlation analysis with other LSS tracers \citep{Koukoufilippas:2020, chiang2020thermal} and  direct observations of clusters \citep{Smith:2016, Eckert:2019}. However, a previous study by \citet{Hill_2014}, analyzing cross-correlations between CMB lensing and Compton-$y$, was sensitive to even higher redshift but reported a mass bias consistent with unity. Note that \citet{Hill_2014} used a slightly different cosmology for their analysis and probed the redshifts that are more impacted by the CIB contamination and its appropriate mitigation strategy. Similarly, an earlier analysis by \citet{Ma_2015} used shear-$y$ correlations and obtained a lower mass bias value, but they also used a slightly different cosmology and ignored the impact of CIB which we find to be significant (see \paperA). We also note that the galaxy cross-correlation analysis of \citep{chiang2020thermal,Koukoufilippas:2020} and $q_{\rm cut}=6$ analysis of \citet{rotti2020removing} are sensitive to lower mass halos compared to our peak sensitivity (see Fig.~\ref{fig:gty_sens}). We defer a detailed analysis of the evolution of mass bias parameter with halo masses to a future study (c.f. \citet{Barnes_2021}). Although the model of redshift evolution of mass bias awaits future data to obtain more precise constraints, this analysis shows how a redshift evolution of sign and magnitude found here can resolve apparent tensions in the inference of this quantity from different probes.

\section{Discussion}
\label{sec:conclusion}

This is the second paper in a series of two on the analysis of the cross-correlation between gravitational lensing shears from DES Y3 data and Compton-$y$ measurements from ACT and $\planck$.  The total signal-to-noise of these measurements is approximately 21, the highest significance measurement of the shear-$y$ correlation to date.  The companion paper \citep{paper1} presented the measurements and systematic tests, and analyzed how well the data fit the feedback predictions from hydrodynamical simulations. In this paper, we take an alternate approach, varying the parameters describing the pressure profiles of halos in our fits to the data.

The shear-$y$ correlation is sensitive to the pressure profiles across a wide range of halo mass and redshift.  Our particular measurements are most sensitive to the pressure within halos with masses of ${\rm few}\times 10^{13}\,M_{\odot} \lesssim M \lesssim 2\times 10^{15}\,M_{\odot}$ and redshifts $z \lesssim 0.8$, as seen in Fig.~\ref{fig:gty_sens}.  We fit the measured shear-$y$ correlation to constrain the redshift and halo mass-dependence of the pressure profiles of dark matter halos.  Our fits are performed at fixed cosmological parameters, but we present results using both the best-fit $\planck$ and best-fit DES-Y1 parameters. Our main results do not depend on this choice, although our quantitative conclusions are somewhat sensitive to the assumed cosmological model.

Our main findings are as follows:
\begin{itemize}
    \item The shear-$y$ correlation measurements are fit reasonably well by a halo model based on the pressure profile of \citet{Battaglia:2012}, but which introduces additional freedom in the mass-dependence of the pressure profile for low-mass ($M < 10^{14}\,M_{\odot}$) halos (Fig.~\ref{fig:residual_besfit})
    \item Our model fits prefer lower amplitude pressure profiles at low halo mass (Fig.~\ref{fig:ym_comp} and Fig.~\ref{fig:pe_3d_comp}) and weakly prefer stronger redshift evolution than predicted by the \citet{Battaglia:2012} model
    \item Our inference of the amplitude of the pressure profiles of low-mass halos is consistent with predictions from hydrodynamical simulations that include the impact of AGN feedback (Fig.~\ref{fig:ym_comp})
    \item Our findings are generally consistent with measurements of the galaxy-$y$ correlation from \citet{Hill:2018} and \citet{Pandey:2019}, and constraints on the $y$ autospectrum from SPT and ACT.
    \item We infer the hydrostatic mass bias from our analysis, finding that its value can change when assuming a lower-$\sigma_8$ than \planck (see Fig.~\ref{fig:hydroMB_comp}). We also find that while assuming a redshift evolution significantly increases the uncertainity on the hydrostatic mass bias, its inferred mean value changes with the correct sign and sufficient magnitude, which can also resolve the apparent tension between this quantity obtained from different probes (see Fig.~\ref{fig:hydroMB_comp_lit}). 
    \item We model the impact of intrinsic alignments on our analysis, finding it to have a small but non-negligible impact. Previous analyses have ignored this effect.
\end{itemize}

The shear-$y$ correlation provides a powerful probe of the thermal energy distribution throughout the Universe. This probe also bridges the gap in the halo-mass sensitivity of galaxy-$y$ correlations and Compton-$y$ auto-correlations. Our measurements suggest that the thermal energy in low-mass halos ($M < 10^{14}\,M_{\odot}$) is suppressed relative to predictions that ignore the impact of AGN feedback. These findings will be crucial in estimating the impact of baryonic physics on cosmological analyses using the cosmic shear data from ongoing and future photometric surveys. We also expect inclusion of kinematic SZ (kSZ) effect and its cross-correlations with tracers of the large scale structure to provide complementary constraints on the physics of feedback (see \cite{Schaan_2021, amodeo2020atacama}). We leave a joint analysis of tSZ and kSZ effects and its cross-correlations with the shear field to a future study. 

Our findings suggest that we will be able to answer important and outstanding questions related to the physics of hot gas and its cosmological implications using the lower noise Compton-$y$ maps covering a larger area from ongoing and future CMB experiments. The imminent release of Compton-$y$ maps from ongoing high resolution surveys like ACT and SPT, as well as future experiments like Simons Observatory\footnote{https://simonsobservatory.org/} and CMB-S4\footnote{https://cmb-s4.org/} would significantly decrease the statistical uncertainty in small scales which are sensitive to smaller mass and higher redshift halos, and are therefore more sensitive to the feedback mechanisms. Moreover, availability of deeper and lower noise shear catalogs from DES in coming years as well as larger scale surveys like the Euclid Space Telescope,\footnote{https://www.euclid-ec.org} the Dark Energy Spectroscopic Instrument,\footnote{https://www.desi.lbl.gov} the Nancy G. Roman Space Telescope,\footnote{https://roman.gsfc.nasa.gov} and the Vera C. Rubin Observatory Legacy Survey of Space and Time\footnote{https://www.lsst.org} will result in a qualitative improvement in the shear-$y$ correlation as a probe, advancing our understanding of feedback physics.

\section*{Acknowledgements}
% \section*{Acknowledgements}
This paper has gone through internal review by the DES and ACT collaborations. SP is supported in part by the US Department of Energy Grant No. DE-SC0007901 and NASA ATP Grant No. NNH17ZDA001N. ES is supported by DOE grant DE-AC02-98CH10886. ES is supported by DOE grant DE-AC02-98CH10886. KM acknowledges support from the National Research Foundation of South Africa. ZX is supported by the Gordon and Betty Moore Foundation. JPH acknowledges funding for SZ cluster studies from NSF AAG number AST-1615657. ADH acknowledges support from the Sutton Family Chair in Science, Christianity and Cultures. CS acknowledges support from the Agencia Nacional de Investigaci\'on y Desarrollo (ANID) under FONDECYT grant no.\ 11191125.

Funding for the DES Projects has been provided by the U.S. Department of Energy, the U.S. National Science Foundation, the Ministry of Science and Education of Spain, 
the Science and Technology Facilities Council of the United Kingdom, the Higher Education Funding Council for England, the National Center for Supercomputing 
Applications at the University of Illinois at Urbana-Champaign, the Kavli Institute of Cosmological Physics at the University of Chicago, 
the Center for Cosmology and Astro-Particle Physics at the Ohio State University,
the Mitchell Institute for Fundamental Physics and Astronomy at Texas A\&M University, Financiadora de Estudos e Projetos, 
Funda{\c c}{\~a}o Carlos Chagas Filho de Amparo {\`a} Pesquisa do Estado do Rio de Janeiro, Conselho Nacional de Desenvolvimento Cient{\'i}fico e Tecnol{\'o}gico and 
the Minist{\'e}rio da Ci{\^e}ncia, Tecnologia e Inova{\c c}{\~a}o, the Deutsche Forschungsgemeinschaft and the Collaborating Institutions in the Dark Energy Survey. 

The Collaborating Institutions are Argonne National Laboratory, the University of California at Santa Cruz, the University of Cambridge, Centro de Investigaciones Energ{\'e}ticas, 
Medioambientales y Tecnol{\'o}gicas-Madrid, the University of Chicago, University College London, the DES-Brazil Consortium, the University of Edinburgh, 
the Eidgen{\"o}ssische Technische Hochschule (ETH) Z{\"u}rich, 
Fermi National Accelerator Laboratory, the University of Illinois at Urbana-Champaign, the Institut de Ci{\`e}ncies de l'Espai (IEEC/CSIC), 
the Institut de F{\'i}sica d'Altes Energies, Lawrence Berkeley National Laboratory, the Ludwig-Maximilians Universit{\"a}t M{\"u}nchen and the associated Excellence Cluster Universe, 
the University of Michigan, the National Optical Astronomy Observatory, the University of Nottingham, The Ohio State University, the University of Pennsylvania, the University of Portsmouth,  SLAC National Accelerator Laboratory, Stanford University, the University of Sussex, Texas A\&M University, and the OzDES Membership Consortium.

Based in part on observations at Cerro Tololo Inter-American Observatory at NSF's NOIRLab (NOIRLab Prop. ID 2012B-0001; PI: J. Frieman), which is managed by the Association of Universities for Research in Astronomy (AURA) under a cooperative agreement with the National Science Foundation.

The DES data management system is supported by the National Science Foundation under Grant Numbers AST-1138766 and AST-1536171. The DES participants from Spanish institutions are partially supported by MINECO under grants AYA2015-71825, ESP2015-66861, FPA2015-68048, SEV-2016-0588, SEV-2016-0597, and MDM-2015-0509, 
some of which include ERDF funds from the European Union. IFAE is partially funded by the CERCA program of the Generalitat de Catalunya. Research leading to these results has received funding from the European Research Council under the European Union's Seventh Framework Program (FP7/2007-2013) including ERC grant agreements 240672, 291329, and 306478. We  acknowledge support from the Brazilian Instituto Nacional de Ci\^encia e Tecnologia (INCT) e-Universe (CNPq grant 465376/2014-2).

This manuscript has been authored by Fermi Research Alliance, LLC under Contract No. DE-AC02-07CH11359 with the U.S. Department of Energy, Office of Science, Office of High Energy Physics.

Support for ACT was through the U.S. National Science Foundation through awards AST-0408698, AST-0965625, and AST-1440226 for the ACT project, as well as awards PHY-0355328, PHY-0855887 and PHY-1214379. Funding was also provided by Princeton University, the University of Pennsylvania, and a Canada Foundation for Innovation (CFI) award to UBC. ACT operates in the Parque Astron'omico Atacama in northern Chile under the auspices of the Agencia Nacional de Investigaci'on y Desarrollo (ANID).The development of multichroic detectors and lenses was supported by NASA grants NNX13AE56G and NNX14AB58G. Detector research at NIST was supported by the NIST Innovations in Measurement Science program.

\bibliography{ref}

\appendix

\section{Covariance matrix}\label{app:cov}
Our full model of theory covariance, including the Gaussian and non-Gaussian terms is shown is Eq.~\ref{eq:cov_tot}. In \paperA $\,$ we validated the Gaussian part of our total covariance using simulations. We have also compared it to the jackknife covariance estimate which partly captures the non-Gaussian contribution to the total covariance. Our total covariance includes the contribution from poisson fluctuations of large clusters. 

\begin{figure*}[h]
\includegraphics[width=1.0\textwidth]{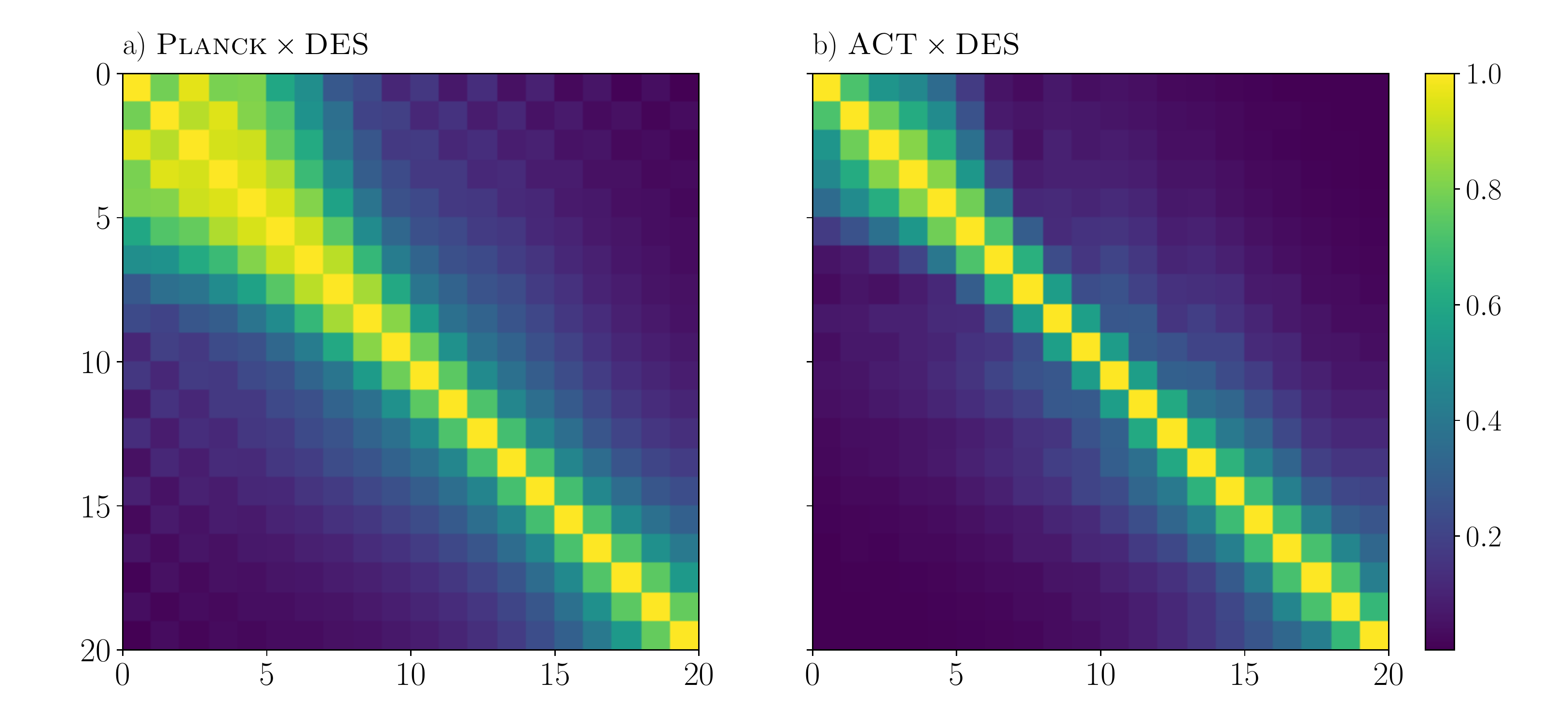}
\caption[]{Correlation matrix of $\xigty$ using the fourth source bin and the two Compton-$y$ maps, binned into 20 radial bins from 2.5 to 250 arcmin.}
\label{fig:corr_mat}
\end{figure*}

In Fig.~\ref{fig:corr_mat} we show the part of the correlation matrix using fourth source tomographic bin. It clearly shows that due to large beam, the small scale angular bins corresponding to $\theta < 10$arcmin are more correlated in the $\planck \times \des$ part of the matrix compared to $\act \times \des$.

\section{Fits with Arnaud10 model}\label{app:fitA10}

\begin{figure*}[h]
	\includegraphics[width=1.0\textwidth]{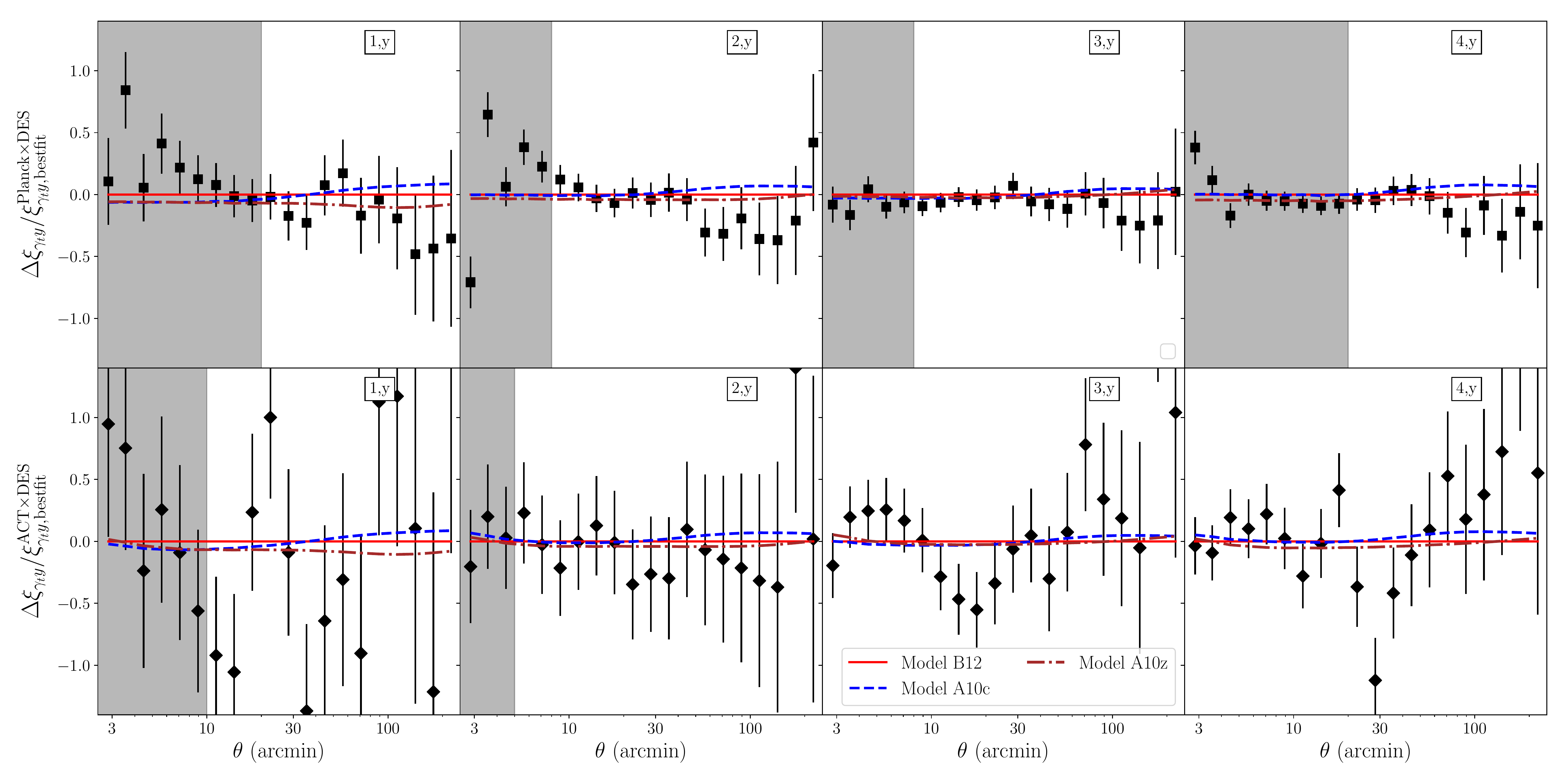}
	\caption[]{This figure is simiar to Fig.~\ref{fig:residual_besfit} but comparing the best fit to the data obtained from the \citet{Arnaud:2010}  model and the \citet{Battaglia:2012} model. We show the best fit for the case of fixing $\rho_B = 0$ (dashed curve, A10c model) and free $\rho_B$ (dot-dashed curve, A10z model). We refer the reader to Table~\ref{tab:params_all} for details of the model parameters and priors used in the analyses.}
	\label{fig:A10_fits}
\end{figure*}

In Fig.~\ref{fig:A10_fits} we compare the best fits obtained from the models based on \citet{Arnaud:2010} with the one obtained from the  \citet{Battaglia:2012} model (as shown in Fig.~\ref{fig:residual_besfit}).  We find that all three models result in similar goodness of fit. The PTE for A10c model is 0.02 and for A10z model is 0.0198. 

\section{Impact of assumed cosmological model on parameter constraints}\label{app:cosmo_comp}

\begin{figure*}[h]
\includegraphics[width=0.95\textwidth]{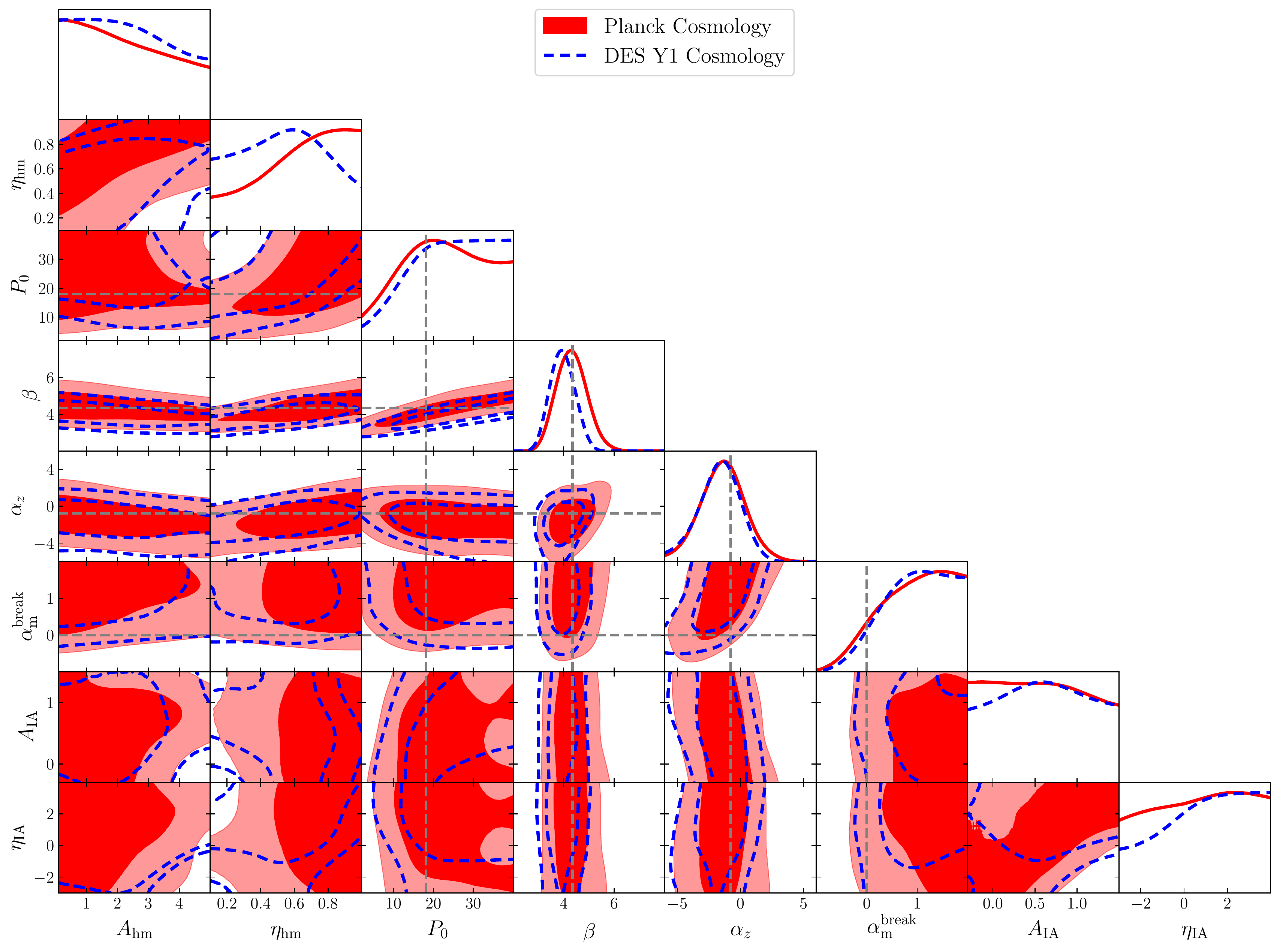}
\caption[]{Constraints on the pressure profile parameters from the \textit{break} model assuming two different cosmological models. }
\label{fig:cosmo_impact}
\end{figure*}

We repeat our analysis adopting the best-fit cosmological parameters from \citet{Planck:2018cosmo} and from the DES Year 1 analysis of \citep{des_y1_3x2pt}.  The full posteriors for these two analyses are shown in Fig.~\ref{fig:cosmo_impact}.  We find that our results are largely insensitive to the choice of cosmological model.

\end{document}